\documentclass[
a4paper,
amsmath,
amssymb,
aps,
prd,
nofootinbib,
preprintnumbers,
longbibliography
]{revtex4-2}

\usepackage{graphicx}
\usepackage{dcolumn}
\usepackage{bm}

\begin{document} 

\preprint{MPP-2023-52}

\title{Observation of the dead cone effect in charm and bottom quark jets and its QCD explanation}

\author{Stefan Kluth}
\author{Wolfgang Ochs}
\affiliation{Max-Planck-Institut f\"ur Physik, F\"ohringer Ring 6, 80805 M\"unchen, Germany}
\author{Redamy Perez Ramos}
\affiliation{DRII-IPSA, Bis, 63 Boulevard de Brandebourg, 94200 Ivry-sur-Seine, France, \\
Laboratoire de Physique Th\'eorique et Hautes Energies (LPTHE), UMR 7589,\\ Sorbonne Universit\'e et CNRS, 4 place Jussieu, 75252 Paris Cedex 05, France}

\begin{abstract}
The production of a heavy quark is accompanied by gluon bremsstrahlung which is suppressed at small angles $\Theta\lesssim M_Q/E$ for mass $M_Q$ and high energy $E$ according to perturbative Quantum Chromo Dynamics (QCD) (``dead cone effect''). As particles at small angles typically have large momenta, the heavy quark mass also causes a suppression of high momentum particles. In this paper, we studied this effect in c- and b-quark events using data from Z boson decays in $e^+e^-$ annihilation. The heavy quark fragmentation function for charged particles is reconstructed in the momentum fraction variable $x$ or $\xi=\ln(1/x)$ by removing the decays of the heavy quark hadrons. Indeed, we find an increasing suppression of particles with rising $x$ down to a fraction of $\lesssim 1/10$ for particles with $x\gtrsim0.2$ in b-quark and $x\gtrsim0.4$ in c-quark jets in comparison to light quark fragmentation. The sensitivity to the dead cone effect in the present momentum analysis is considerably increased in comparison to the recently presented angular analysis. This amount of suppression and the differences between c- and b-quark fragmentation are in good quantitative agreement with the expectations based on perturbative QCD within the Modified Leading Logarithmic Approximation (MLLA) in the central kinematic region. The data also support a two parameter description in the MLLA of these phenomena (``Limiting Spectrum''). The sensitivity of these measurements to the heavy quark mass is investigated.
\end{abstract}

\maketitle

\section{Introduction}
\label{sec:intro}

The dead cone effect is a prediction of QCD, the theory of strong interactions within the Standard Model of particle physics. It originates from the radiation pattern off a heavy quark as obtained in perturbation theory~\cite{Dokshitzer:1991fc,Dokshitzer:1991fd}.
For an energetic heavy quark $Q$ of mass $M_Q$ and energy $E_Q$ such that $E_Q/M_Q \gg 1$, the gluon emission probability for small emission angle $\Theta$ and low energy $\omega$, can be written as
\begin{equation}
  d\sigma_{Q\to Q+g} \simeq \frac{\alpha_s}{\pi}C_F\frac{\Theta^2d\Theta^2}{(\Theta^2+\Theta^2_0)^2} \frac{d\omega}{\omega},
\label{emission}
\end{equation}
with angular cut-off $\Theta_0=M_Q/E_Q$; $\alpha_s$ denotes the strong coupling constant and $C_F$ the QCD colour factor at the branching vertex $Q\to Q+g$. Therefore, for smaller emission angles $\Theta<\Theta_0$, gluon radiation is suppressed and vanishes in the forward direction such that the region with the gluon depopulated cone around the flight direction of the heavy quark $Q$ is called ``dead cone''. For large emission angles $\Theta\gg \Theta_0$, the gluon radiation pattern becomes identical to that of a light quark jet, and the same statement holds for the internal angular ordered structure of secondary gluon subjets.

In the early studies, as a first consequence of the dead cone effect, a reduction of the full particle multiplicity in the heavy quark jet has been predicted. This effect has indeed been observed in~\cite{Schumm:1992xt} and in the subsequent update~\cite{Dokshitzer:2005ri} with results nearby the QCD expectation. Only recently, a more direct observation of the dead cone effect has been achieved by the ALICE collaboration~\cite{ALICE:2021aqk}, which has presented results on the differential angular structure of charm-quark and inclusive jets from proton-proton collisions at $\sqrt{s}=13$~TeV at the Large Hadron Collider (LHC). A relative suppression of small angle particle emission is observed in the heavy quark jet in agreement with Monte Carlo Event Generators (MCEG) combining the hard interactions of the partons from the protons with a QCD parton shower and a hadronisation model. Already before, some preliminary angular studies of the dead cone effect have been presented using data on charm-jets from HERA~\cite{Perieanu:2006vn} and data on b-jets from LEP~\cite{Battaglia:2004coa}.

Multi-parton final states in quark and gluon jets can be calculated perturbatively from subsequent parton branchings with running coupling $\alpha_s$ down to the transverse momentum cut-off $Q_0$, within the probabilistic parton shower picture based on angular ordering~\cite{Bassetto:1983mvz,Fadin:1983aw,Marchesini:1983bm,DKMT-book}. Some insight can be gained by first considering the Double Logarithmic Approximation (DLA), which accounts for leading collinear and soft singularities as in eq.~\eqref{emission} for $\Theta_0=0$. A more accurate description is achieved in the MLLA if single logarithmic terms are included (relative order ${\cal O}(\sqrt{\alpha_s})$).
In the comparison with experiments, the hypothesis of ``Local Parton Hadron Duality'' (LPHD)~\cite{Azimov:1984np} has turned out to be quite successful in many applications. In this scheme, perturbative QCD predictions on sufficiently inclusive observables for final state partons are in close correspondence with these observables for hadrons, where $Q_0$ may achieve low values, even down to the mass scale $\Lambda$ of QCD (``Limiting Spectrum'').

The single inclusive gluon spectrum for a quark or gluon jet in the variable $\xi=\ln(1/x)$, with $x=E_{g}/E$, has been computed as function of the primary energy $E$ in MLLA~\cite{Azimov:1984np,Azimov:1985by,Fong:1990nt}.
The main feature is a Gaussian-like peak in the $\xi$ spectrum, the so-called ``hump-backed plateau'' resulting from the coherent emission of soft gluons in the cascade. Predictions for the $\xi$ spectrum and the primary energy dependence of the Gaussian parameters agree well with $e^+e^-$ annihilation data, see e.g.~\cite{Kluth:2006bw} for a review and~\cite{Perez-Ramos:2013eba} for a recent analysis. Generally, only charged particles are studied experimentally since the resolution for direction and momentum measurements is improved with respect to neutral particles, in particular for low momentum particles dominating the hadronic final states. For an overview of the perturbative QCD approach on parton shower evolution, approximations and some applications to multiparticle production, see e.g.~\cite{DKMT-book,Khoze:1996dn}.

In this paper the dead cone effect is studied for the first time by exploring the internal momentum structure of heavy and light-quark jets, which can be directly compared with the early QCD predictions within the MLLA-LPHD approach~\cite{Dokshitzer:1991fc,Dokshitzer:1991fd}. In jets particles with large momenta are emitted on average at small angles and particles with small momenta at larger angles. Therefore, the dead cone effect at small angles corresponds to a suppression of large momentum gluons. An advantage of studying momentum spectra is that their determination does not require a jet axis definition which may cause important systematic uncertainties.

\section{Experimental determination of particle spectra in $b$ - and $c$ - quark events}
\label{sec:experiment}

\subsection{Reconstruction of heavy quark fragmentation functions}
\label{sec:Reconstruction}

Experimental data are presented as function of the charged particle momenta $p$, with $x_p=2p/W$ or $\xi_p=\ln (1/x_p)$ at c.m.s.\ energy $W$. In $e^+e^-$ annihilation, we refer to the fragmentation function 
\begin{equation}\label{eq:FF}
  \bar{D}(\xi,W)=\frac{1}{2}\frac{1}{\sigma_{tot}} \frac{d\sigma^h}{d\xi} (\xi,W)
\end{equation}
where $\bar D(\xi,W)=x D(x,W)$ with the inclusive $x$-distribution $D(x,W)$ for particles from both hemispheres. In order to obtain the heavy quark fragmentation function $\bar{D}_Q^{ch}(\xi_p,W)$, $Q=c,b$, we start from the measured $\xi_p$-distribution of light hadrons in events tagged as originating from $Z\rightarrow Q{\bar Q}$ decays. This distribution also contains the charged hadrons from B-hadron or Charm-hadron decays and they have to be subtracted. The $\xi_p$-distributions of charged B-hadron or Charm-hadron decay products have not been measured separately\footnote{A measurement could use the impact parameters of reconstructed tracks w.r.t.\ the primary vertex to select the heavy hadron decay products.} and we obtain them from a MCEG program; the most recent version of Pythia8~\cite{Bierlich:2022pfr,Skands:2014pea} is used for this purpose with 100'000 events generated. 

\begin{figure}[htpb!]
\begin{tabular}{cc}
\hskip -1.1cm
\includegraphics[height=8.5cm,width=9.5cm]{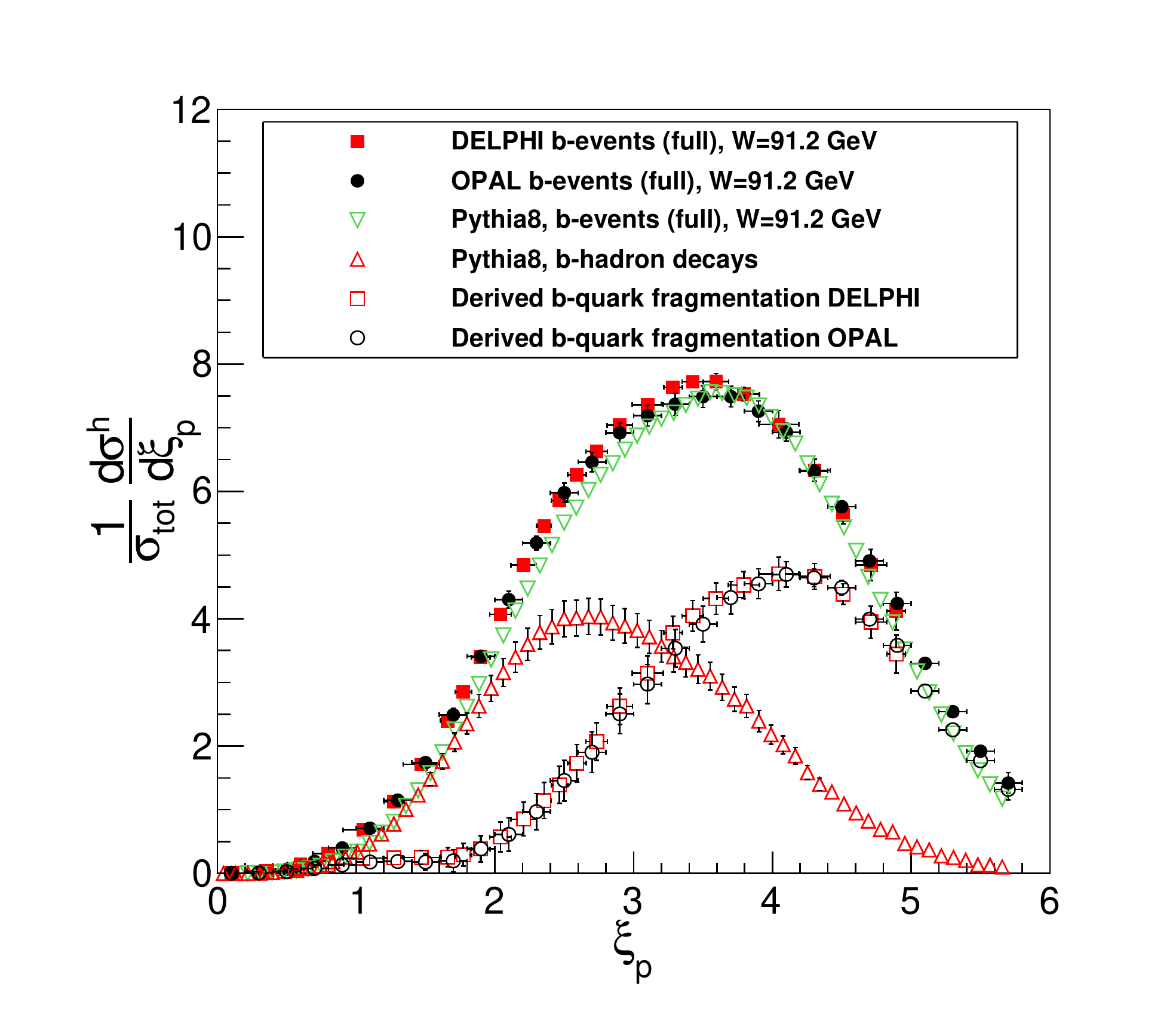} &
\hskip -1cm
\includegraphics[height=8.5cm,width=9.5cm]{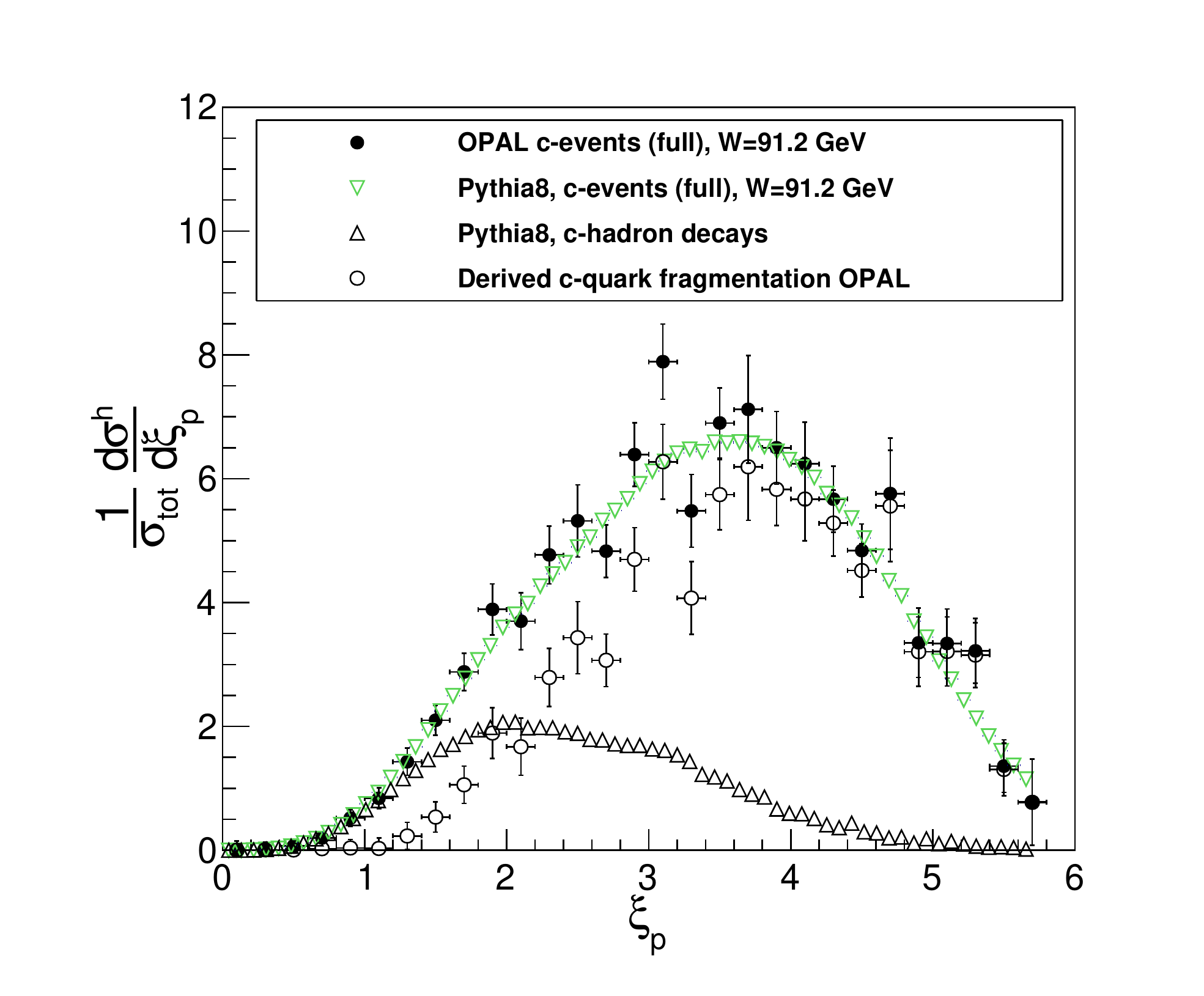} \\
\end{tabular}
\caption{Distribution of charged hadrons in $\xi_p= \ln (1/x_p)$ for full $b\bar b$ events including B-hadron decays at 91.2~GeV: DELPHI and OPAL data and prediction from the Pythia8 MCEG; B-hadron decay charged particle $\xi_p$-distribution from Pythia8 with systematic error ($6.5\%$); the b-quark fragmentation functions as derived from DELPHI and OPAL data after subtraction of the B-hadron decay charged particle distribution rescaled to the experimental decay multiplicity eq.~\eqref{ndecay} (left panel); corresponding results for $c\bar c$ events: full spectrum by OPAL and from Pythia8, Charm-hadron decay charged particle distribution from Pythia8 and the c-quark fragmentation function (right panel).}
\label{fig:decays}
\end{figure}

The full $\xi_p$-distribution of events with B-hadrons including their decays has been measured by several groups~\cite{DELPHI:1998cgx,OPAL:1998arz,L3:2004cdh,ALEPH:1995njx}
with good mutual agreement. The results obtained\footnote{Only DELPHI and OPAL presented their data for b- or c-quarks and uds-quarks in $\xi$.} by DELPHI~\cite{DELPHI:1998cgx} and OPAL~\cite{OPAL:1998arz} are shown in Fig.~\ref{fig:decays} together with the MCEG prediction for this distribution (left panel). The predictions agree within $\sim$$10\%$ with the data. The measured charged hadron multiplicities $N_{b}^{ch}=23.17\pm0.35$ (DELPHI) and $N_{b}^{ch}=23.16\pm0.45$ (OPAL) compare to $N_{b}^{ch}=21.26\pm0.02$, as obtained with Pythia8.
Furthermore in Fig.~\ref{fig:decays}, the $\xi_p$-distribution of the charged B-hadron decay particles is displayed which integrates up to the multiplicity $n^{dec}_{b}=9.80$.

The parameters of the MCEG program are tuned to a large variety of data from $e^+e^-$ and pp collisions~\cite{Skands:2014pea}. Therefore, one cannot expect an optimal agreement in all processes. Indeed, for the B-hadron decay multiplicity, an experimental value has been determined with a rather small error~\cite{Dokshitzer:2005ri} (practically identical to the earlier result~\cite{Schumm:1992xt}) which differs from the MCEG prediction:
\begin{equation}
  n^{dec}_{b}=9.80 \ \  (\rm{Pythia8})\ \ \ \ \ \ n^{dec}_{b}=11.10\pm0.18,\ \  (\rm{experiment}).
\label{ndecay}
\end{equation} 
The experimental result is based on the evaluation~\cite{Bdecaymult} of the measurements by ALEPH, CDF, DELPHI, L3, OPAL and SLD for the single B-hadron decay multiplicity $N_{b}= 4.955\pm0.062$ and includes contributions from $K^0$ and $\Lambda$ decays among others. This number for $n^{dec}_b$ exceeds the one obtained by Pythia8 by $13\%$.

In our subsequent analysis, only the shape of the $\xi_p$-distribution is taken from the MCEG simulation but its normalisation is scaled by $13\%$ to obtain the experimental decay multiplicity eq.~\eqref{ndecay}. We add a systematic error of $6.5\%$ to all data points in order to allow for variations in a band of the missing decay rate.  
Subtracting this rescaled B-hadron decay distribution from the spectra of the full b-events by DELPHI and OPAL, the final b-quark fragmentation function is derived and displayed in Fig.~\ref{fig:decays} (left panel) as well. There is a good agreement between the results from both experiments. 
We have compared our result for the B-hadron decay distribution with the results obtained by the DELPHI collaboration~\cite{DELPHI:1998cgx} using the JETSET MCEG by adding the contributions from $\pi,K,p$.
The b-quark fragmentation function computed with their result agrees with ours within the errors, lying for $\xi_p\gtrsim 3$ below our result at the edge of the error bars. 

The corresponding results for $c\bar c$-events are shown in Fig.~\ref{fig:decays} (right panel). The $\xi_p$-distribution for the full events including the Charm-hadron decays as obtained by OPAL~\cite{OPAL:1998arz} are found in good agreement with the Pythia8 results: the full multiplicity by OPAL $N_c^{ch}= 21.55\pm0.74$ compares with $N_c^{ch}= 20.05$. Also shown is the $\xi_p$-distribution of the charged decay products of the Charm-hadrons which integrates to the multiplicity $n_c^{dec}=5.04$ and compares well with the observed decay multiplicity, $n_c^{dec}=5.2\pm0.3$~\cite{Dokshitzer:2005ri} such that no rescaling is applied. Subtracting the distribution of decay products from the full distribution finally yields the experimentally derived c-quark fragmentation function.

\begin{figure}[htpb!]
\begin{tabular}{cc}
\hskip -1.1cm
\includegraphics[height=10.5cm,width=9.5cm]{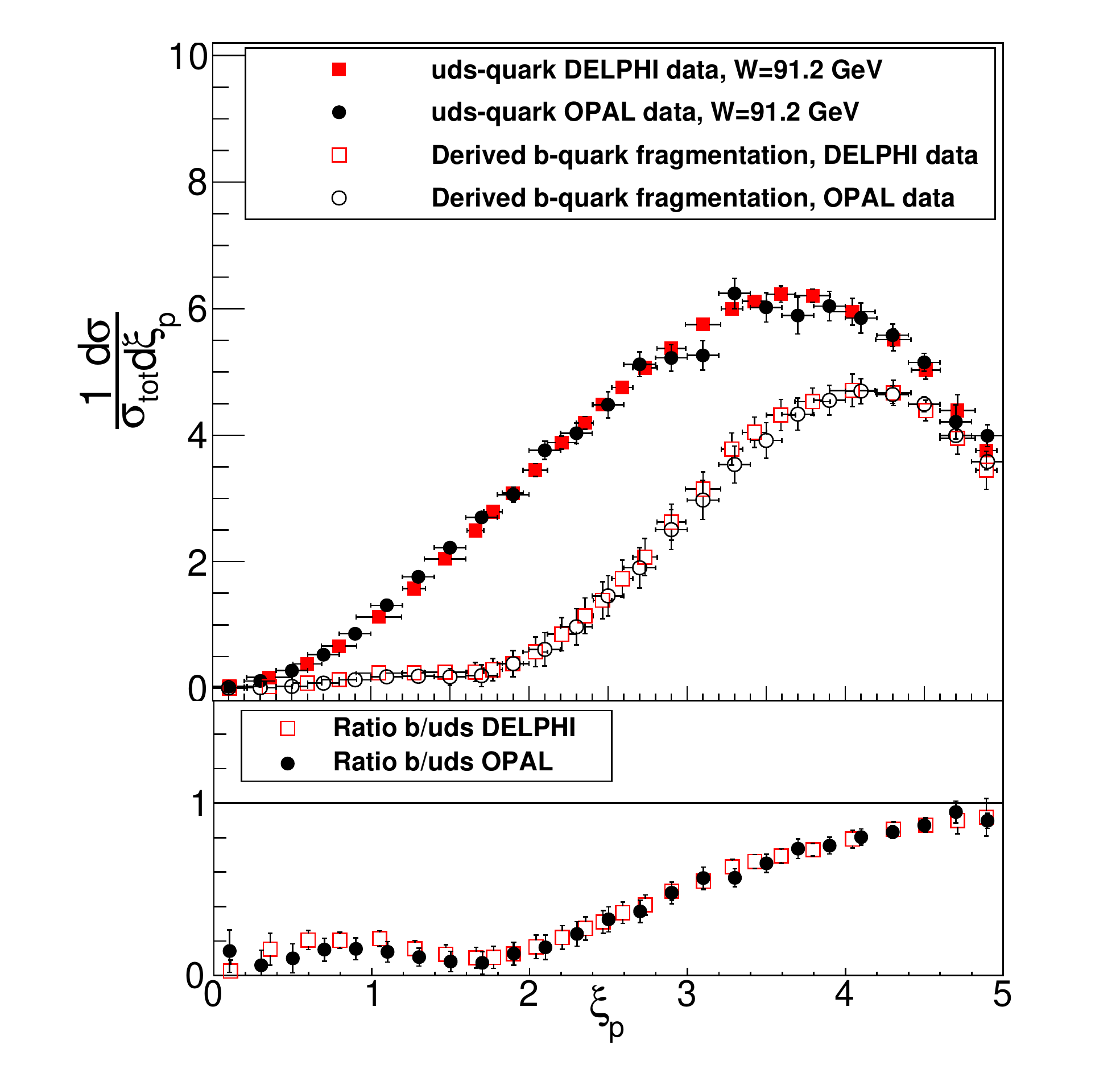} &
\hskip -1cm
\includegraphics[height=10.5cm,width=9.5cm]{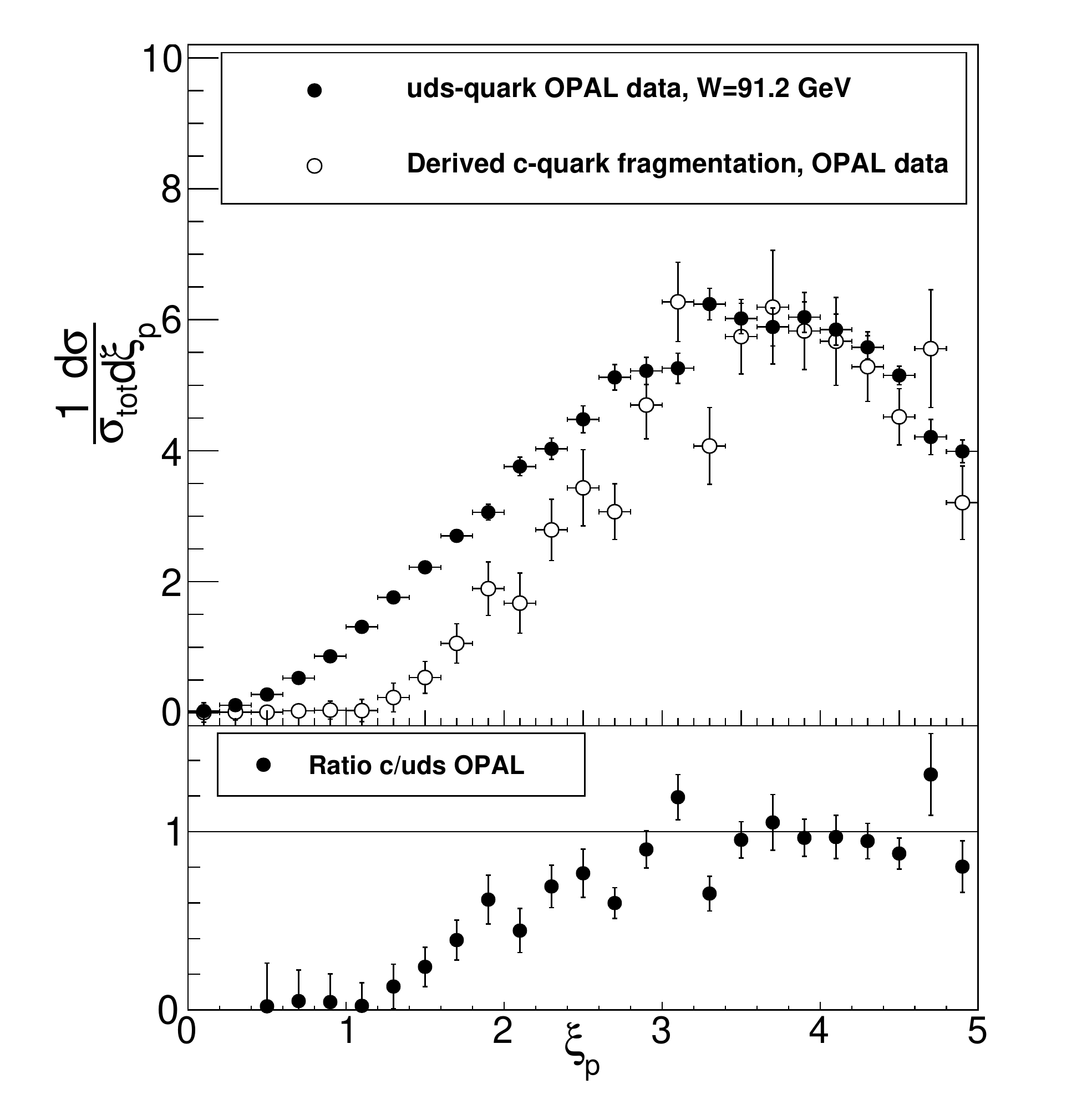} \\
\end{tabular}
\caption{Fragmentation function in $\xi_p= \ln (1/x_p)$ for the light uds-quarks in comparison with experimentally derived $\xi_p$-distributions for the b-quark and c-quark fragmentation (upper panels); ratio of the heavy quark over the light quark fragmentation functions showing the strong suppression of the heavy quark fragmentation for small $\xi_p$ (large momenta) by an order of magnitude which constitutes the dead cone effect (lower panels).}
\label{fig:exp-deadcone}
\end{figure}

\subsection{Evidence for the dead cone effect in heavy quark events}
\label{sec:deadcone}

Next we compare in Fig.~\ref{fig:exp-deadcone} the light uds-quark $\xi_p$-distributions measured by DELPHI and OPAL with the b-quark and c-quark $\xi_p$-distributions derived in the last subsection by removing the charged heavy hadron decay products. One observes the strong suppression of particle production in the b-quark and c-quark fragmentation (upper panels) and in the ratios of both distributions (lower panels). While the ratios approach unity for large $\xi_p$ in the low momentum limit, the suppression progresses down to a ratio of $\lesssim 1/10$ for small $\xi_p$ values (large momenta) before it levels off. There is a notable difference between c- and b-quark fragmentation: the ratio for the b-quark starts decreasing already at $\xi_p\sim 5$ and falls down to $\sim 1/10$ near $\xi_p\sim 2$, whereas for the c-quark, it starts decreasing later at $\xi_p\sim 3$ and falls to zero near $\xi_p\sim 1$. This difference will be related to the different quark masses $M_Q$ below.

The dead cone effect is established in this process with a high significance, which is comfortably larger than five $\sigma$ for both b-quarks and c-quarks.
According to eq.~\eqref{emission}, the dead cone effect is characterised by the full suppression of the small angle fragmentation from the heavy quark. The results in Fig.~\ref{fig:exp-deadcone} show the efficiency of the momentum space analysis in reflecting the almost complete suppression of heavy quark fragmentation in the corresponding limit of large momenta. This suppression is stronger than the maximal suppression by a factor of about $1/2$ observed in the angular analysis by ALICE~\cite{ALICE:2021aqk}. This may be related to the finite jet resolution and the difficulty to define the gluon emission angle in this analysis. 

\section{MLLA expectations for heavy quark fragmentation functions}
\label{sec:theory}

\subsection{The MLLA QCD relations between light and heavy quarks jets}

The dead cone effect has first been studied for total multiplicities of light hadrons in QCD jets within the MLLA of perturbative QCD~\cite{Dokshitzer:1991fc,Dokshitzer:1991fd} and the predictions have been compared with data in~\cite{Schumm:1992xt,Dokshitzer:2005ri}.
In the MLLA the accompanying multiplicity in the production of a heavy quark pair $N_{Q\bar Q} (W)$ at c.m.s.\ energy $W$ can be expressed in terms of the multiplicity $N_{q\bar q} (W)$ in the light quark $q\bar q$ production (q=u,d,s) with the multiplicity in the dead cone subtracted as
\begin{equation}
  N_{Q\bar Q} (W) =  N_{q\bar q} (W)- N_{q\bar q}(\sqrt{e}M_Q), 
\label{dc-multiplicity}
\end{equation}
with the dead cone mass scale $W_0=\sqrt{e}M_Q$ ($e=\rm{exp}(1)$). The directly observed full charged hadron multiplicity in heavy quark events produced in $e^+e^-$ annihilation $N_Q^{ch}\equiv N^{ch}_{e^+e^-\to Q\bar Q}$ can be written as
\begin{equation}
  N_Q^{ch}(W)=  N_{Q\bar Q}^{ch} (W) + n_Q^{dec},
\label{QQbar-multiplicity}
\end{equation}
i.e.\ as the sum of multiplicities accompanying the heavy quarks $N_{Q\bar Q}^{ch}$ and the charged multiplicities from the decays of the two heavy hadrons $n_Q^{dec}$.  

As an important consequence of eq.~\eqref{dc-multiplicity}, the difference between the observed charged particle multiplicities in heavy and light quark events
\begin{equation}
  \delta_{Q\ell}= N_{Q}^{ch}(W) - N_{q\bar q}^{ch}(W)
\label{delta-b-ell-def}
\end{equation}
is predicted in the MLLA as 
\begin{equation}
  \delta_{Q\ell}^{MLLA} = n_Q^{dec} - N_{q\bar q}(\sqrt{e}M_Q),
\label{delta-b-ell}
\end{equation}
such that this quantity is independent of the total energy $W$ and depends only on the heavy quark mass $M_Q$. As reviewed in the dead cone analysis~\cite{Dokshitzer:2005ri}, this difference is indeed found to be independent of the c.m.s.\ energy in $e^+e^-$ annihilation to b-quarks up to LEP~2 energies within the experimental uncertainties. An alternative model without the dead cone effect and a pronounced energy dependence of $\delta_{b\ell}$ has been excluded at high confidence level.

For our analysis of the inclusive spectra, the relation for multiplicities as their integrals, will serve as an important cross check. For the b-quark with the values $n_b^{dec}=11.10\pm 0.18$ and $W_0=8$~GeV with $N_{q\bar q}(8\ \rm{ GeV})=6.7\pm 0.34$, the prediction is close to the experimental result~\cite{Dokshitzer:2005ri}
\begin{equation}
  \delta_{b\ell}^{MLLA} = 4.4\pm0.4, \ \ \  \delta_{b\ell}^{exp}=3.14\pm0.14,
\label{delta}
\end{equation}
but a significant difference remains.
 
For the c-quark, with the values $n_c^{dec}=5.2\pm 0.3$, $W_0=2.7$~GeV and $N_{q\bar q}(2.7\ \rm{GeV})=3.7\pm 0.3$, one finds the charged particle multiplicity difference~\cite{Dokshitzer:2005ri} as 
\begin{equation}
  \delta_{c\ell}^{MLLA} = 1.5\pm0.4, \ \ \  \delta_{c\ell}^{exp}=1.0\pm0.4,
\label{deltac}
\end{equation}
with consistent results between theory and experiment. 

The spectrum $D_Q(x,E)$ of gluons with energy fraction\footnote{For the massless partons $E=p$ in the calculations, in the comparison with light hadrons we take $x=p_h/E$} $x=E_g/E$ at primary energy $E$ accompanying the $Q{\bar Q}$ pair can be treated in a similar way to that of mean multiplicities in eq.~\eqref{dc-multiplicity}; for a review, see~\cite{Khoze:2001aa}. Some insight can be gained by first considering the results for the leading double logarithmic approximation (DLA). The difference between the heavy quark $D_Q(x,E)$ and the light quark $D_q(x,E)$ spectra due to the dead cone effect comes from the radiation of very energetic gluons at small angles $\Theta < \Theta_0$. This radiation can be considered as resulting from a Lorentz boost by the factor $\gamma= E/M_Q$ along the heavy quark direction from the corresponding radiation at lower hardness $M_Q$. In the DLA, a simple formula in analogy to the equation for multiplicities eq.~\eqref{dc-multiplicity} can be written as~\cite{Dokshitzer:1991fd,Dokshitzer:1987aa}: 
\begin{equation}
  D_Q(x,E) = D_q(x,E) -  D_q(x,M_Q).
\label{DLAeq}
\end{equation}

Hence, the heavy quark fragmentation function is represented in terms of the light quark fragmentation functions at the different energy scales $E$ and $M_Q$. This equation cannot be strictly correct, since the $x$-distribution at large $x$ decreases with rising energy because of scale breaking effects, therefore $D_q(x,E) < D_q(x,M_Q)$ and $D_Q(x,E)$ in eq.~\eqref{DLAeq} would become negative (see e.g.\ Fig.~\ref{fig:theoretical_curve} below).

While the equation for multiplicities eq.~\eqref{dc-multiplicity} is derived systematically within the MLLA, the corresponding analysis for the inclusive spectra in MLLA is not yet available at the same rigor. An improved equation for the inclusive $x$-spectra has been presented which reproduces the equation for multiplicities in MLLA after integration over $x$ and avoids a negative fragmentation function. The MLLA estimate has been reported as~\cite{Dokshitzer:1991fd}
\begin{equation}
  \bar D_Q(x,W)= \bar D_q(x,W) -  \bar D_q\left(\frac{x}{\langle x_Q\rangle},\sqrt{e}M_Q\right), 
\label{MLLAeq}
\end{equation}
where $\bar D(x,W)=x D(x,W)$. 
This expression, after integration over the variable $x$, reproduces eq.~\eqref{dc-multiplicity} for the multiplicities. For our comparison with the heavy quark fragmentation function $\bar D_Q(\xi,W)$ with variable $\xi$ as determined in the last section, we rewrite this relation as
\begin{equation}
  \bar D_Q(\xi,W)=\bar D_q(\xi,W) -  \bar D_q( \xi - \xi_Q,\sqrt{e}M_Q),
\label{MLLAeqxi}
\end{equation}
with $\bar D(\xi,W)=x D(x,W)$ and $\xi_Q=\ln(1/\langle x_Q\rangle)$. Again, as in the equation for multiplicities, the low energy scale is $W_0=\sqrt{e} M_Q$. Furthermore, the mean momentum fraction ${\langle x_Q\rangle}$ of the primary heavy quark $Q$ is introduced which reduces the light particle energies to $x < {\langle x_Q\rangle}$. The shift of the $\xi$-spectrum by $\xi_Q$ corresponds to an MLLA correction of ${\cal O}(\sqrt{\alpha_s})$ as can be seen by a Taylor expansion of $\bar D_q(\xi-\xi_Q,\sqrt{e}M_Q)$ in $\xi_Q$ at  $\xi_Q=0$. The eq.~\eqref{MLLAeqxi} represents an approximation that does not work well at small $\xi$ since the shifted contribution $\bar D_q(\xi-\xi_Q,W_0)$ has to vanish for $\xi<\xi_Q$. Comparisons of these predictions with experiment should take these limitations into account.

If ${\langle x_Q\rangle}$ is taken from experiment, the heavy quark fragmentation function at c.m.s.\ energy $W$ can be obtained by the relation eq.~\eqref{MLLAeqxi} from the light quark fragmentation functions at energies $W$ and $W_0$ in absolute normalisation.

As numerical values of these parameters, we take for b-quarks, $W_0=8.0$~GeV\footnote{This value corresponds to a b-quark pole mass $M_b=4.85\pm0.15$ which is consistent with the most recent world average pole mass $M_b=4.78\pm0.06$~\cite{ParticleDataGroup:2022pth}}~\cite{Dokshitzer:2005ri} and the experimental evaluation $\langle x_b \rangle=0.7092\pm0.0025$~\cite{DELPHI:2011aa}. For c-quarks, we use $W_0=2.7$~GeV~\cite{Dokshitzer:2005ri} and the experimental value $\langle x_c\rangle=0.495\pm0.006$~\cite{Baines:2006uw}. This yields the shift parameters
\begin{equation}
  \xi_c=0.70, \ \ \  \xi_b=0.36.
\label{xicbvalues}
\end{equation}
These numbers are also consistent with the results in~\cite{Khoze:1996dn}, based on calculations for the heavy quark $x$-spectra in~\cite{Dokshitzer:1995ev}.

\subsection{Experimental test of the MLLA relation between light and heavy quark jets}
\label{sec:distortedgaussion}

\begin{figure}[bthp!]
\begin{tabular}{cc}
\hskip -1.1cm
\includegraphics[height=8.5truecm,width=9.5truecm]{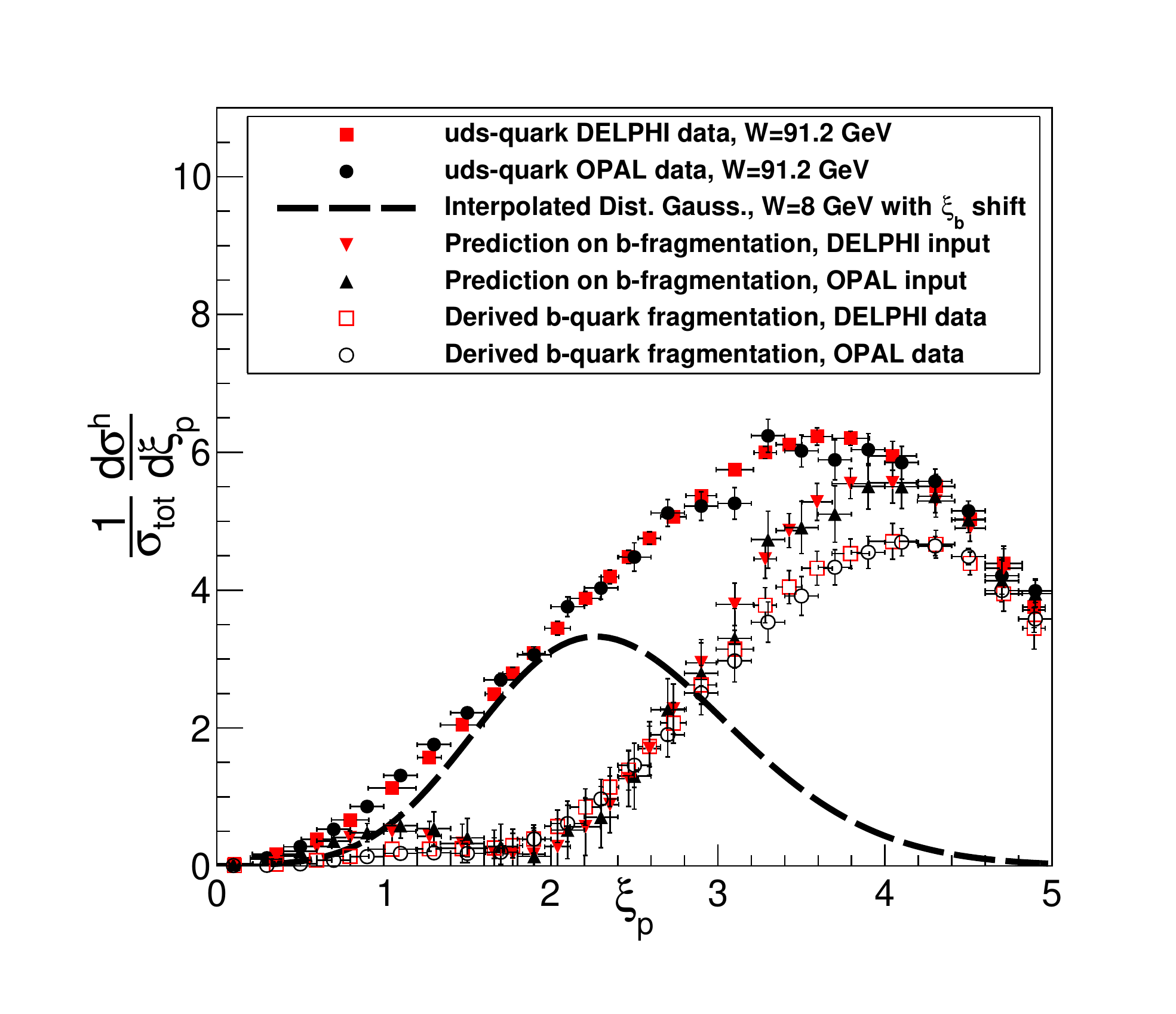} &
\hskip -1.0cm
\includegraphics[height=8.5truecm,width=9.5truecm]{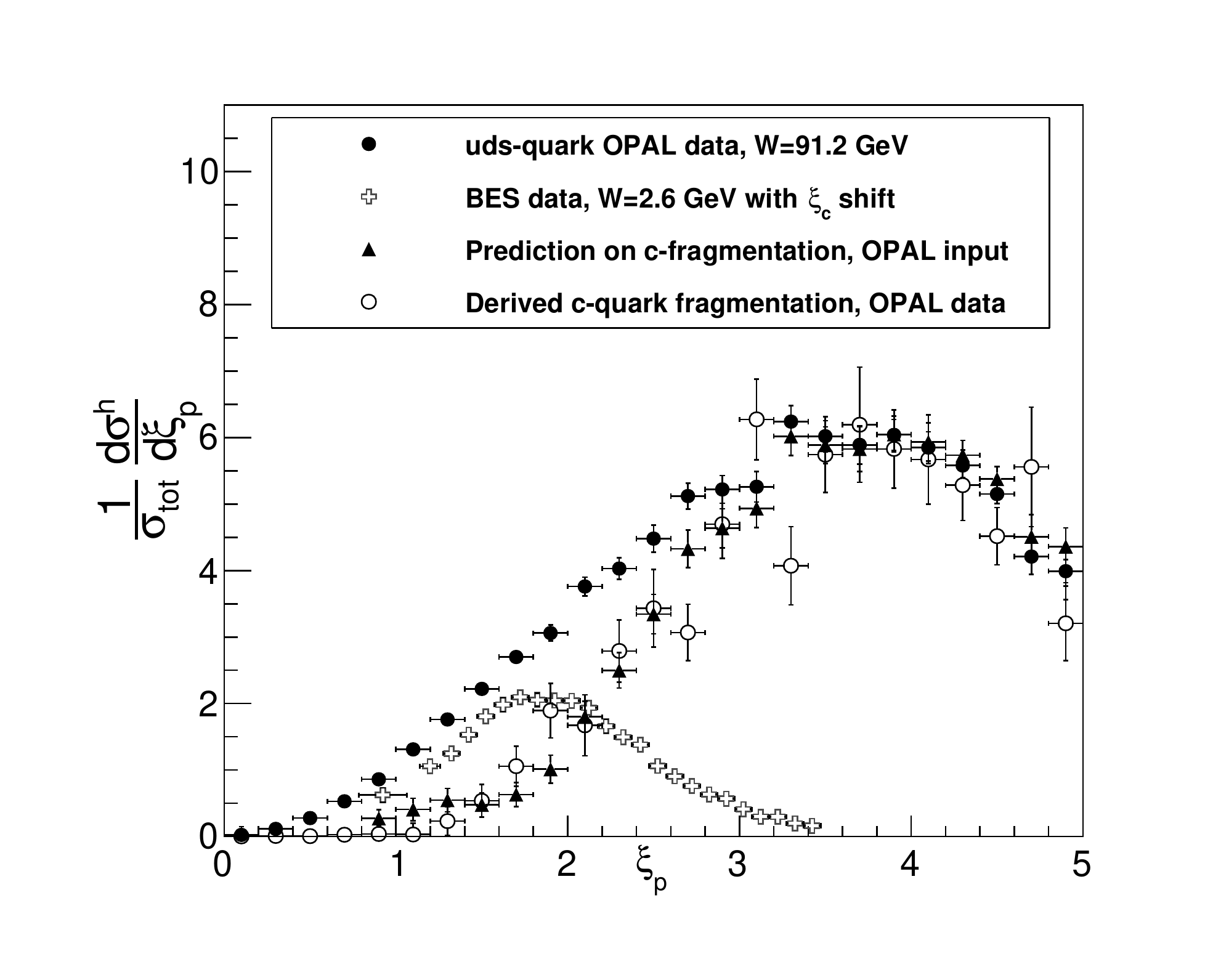} \\
\hskip -1.1cm
\includegraphics[height=8.5truecm,width=9.5truecm]{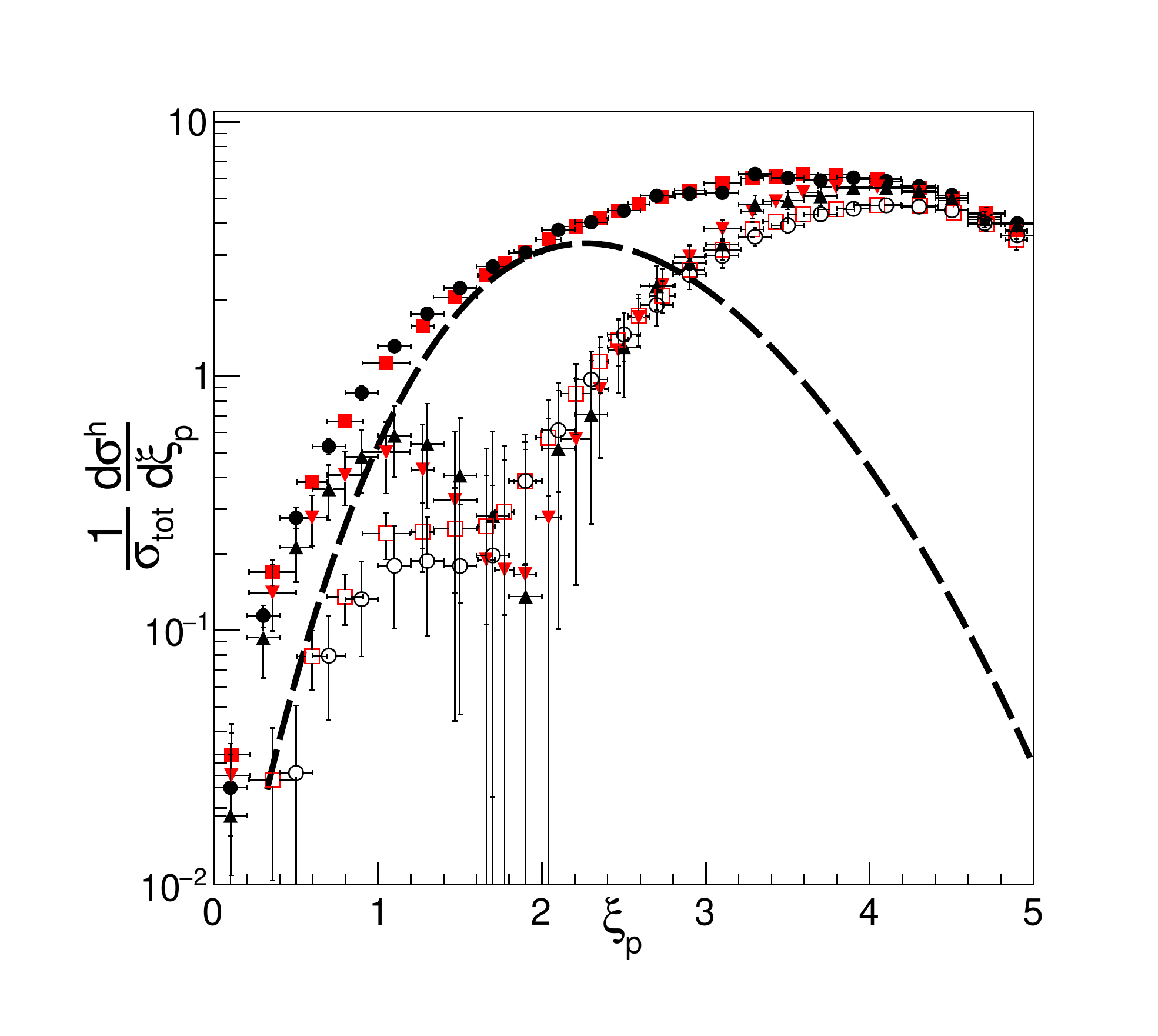} &
\hskip -1.0cm
\includegraphics[height=8.5truecm,width=9.5truecm]{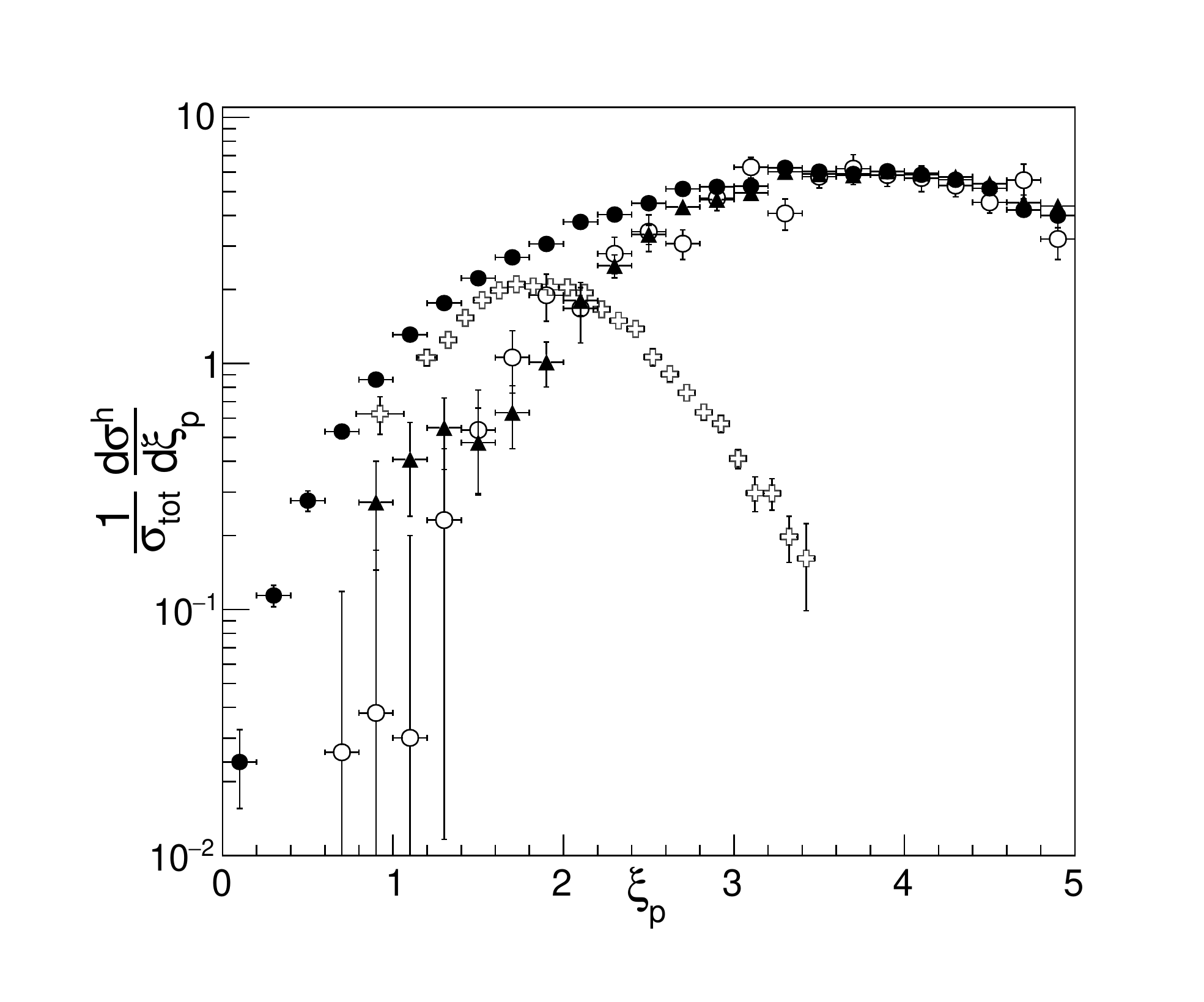} \\
\end{tabular}
\caption{Fragmentation functions in $\xi_p= \ln (1/x_p)$ for uds-quarks at 91.2~GeV and at $W_0=8.0$~GeV by interpolation between neighbouring energies, with correction for charm decays (see App.~\ref{sec:interpolation}) and with shift $\xi_b=0.36$, see eq.~\eqref{xicbvalues}. Subtracting these distributions as in eq.~\eqref{MLLAeqxi}, yields the MLLA prediction for the b-quark fragmentation function to be compared with the experimentally derived one on linear and logarithmic scales (left panels); corresponding results for c-quark fragmentation: at 91.2~GeV the OPAL uds-quark data; at $W_0\sim 2.6$~GeV the data by BES; the MLLA prediction for the c-quark fragmentation function to be compared with the experimentally derived c-quark fragmentation function (right panels).}
\label{fig:ion as displayed in Fig. 1 (left panel) and the same exp_subtraction}
\end{figure}

At first, we probe the MLLA expectation eq.~\eqref{MLLAeqxi} by inserting for $\bar D(\xi,W)$ the experimentally observed distributions in $\xi_p=\ln{1/x_p}$ at the respective energies $W$.
At the low energy $W_0=2.7$~GeV for the c-quark fragmentation, we insert the $\xi_p$-distribution data obtained by the BES collaboration at the nearby energy 2.6~GeV~\cite{BES:2003xdf}. There are no data nearby $W_0=8.0$ GeV for the b-quark fragmentation and therefore, we obtain the corresponding $\xi_p$-distribution from the interpolation between two neighbouring energies, also a correction for charm production has been applied (see App.~\ref{sec:interpolation}). The $\xi_p$-distribution at $W_0=8.0$~GeV, so obtained and shifted by $\xi_b=0.36$ according to eq.~\eqref{MLLAeqxi}, i.e.\ $\bar D_q(\xi-\xi_b,W_0)$, is shown in Fig.~\ref{fig:ion as displayed in Fig. 1 (left panel) and the same exp_subtraction} (left panel) as a dashed line; also shown are the data for $\bar D_q (\xi,W)$ at $W=91.2$~GeV, both referring to uds-quark events. From their difference, according to eq.~\eqref{MLLAeqxi}, one obtains the MLLA prediction for the b-quark distributions $\bar D_b(\xi_p,W)$ where the error bars shown include the systematic errors.  

This MLLA-expected distribution for the b-quark is now compared in Fig.~\ref{fig:ion as displayed in Fig. 1 (left panel) and the same exp_subtraction} to the experimentally derived b-quark fragmentation functions using DELPHI and OPAL data (see Fig.~\ref{fig:decays} (left panel) as discussed in the last section). Again we point to the strong suppression of the b-quark fragmentation function which becomes almost complete for the high momentum particles with $\xi_p\lesssim2$. The MLLA expectations match with the experimentally derived b-quark fragmentation data at a quantitative level in the region around the peak of the 8~GeV distribution between $\xi_p=1.5$ and $\xi_p=3.2$. There are deviations at small $\xi_p$ (large momenta) at a low level of particle density. A larger deviation occurs in the region above $\xi_p=3$, which corresponds to the ultrasoft particles with momenta $p \lesssim Q_0 \sim \Lambda$ at the hadronic mass scale. This region is outside the range of validity of the perturbative approach. 

Now we turn to the results on the c-quark events in the right panel of Fig.~\ref{fig:ion as displayed in Fig. 1 (left panel) and the same exp_subtraction}. By subtracting the low energy BES $\xi_p$-distribution at $W_0=2.6$~GeV, after a shift by $\xi_c= 0.7$, from the uds-quark distribution at $W=91.2$~GeV, one obtains the MLLA-predicted c-quark fragmentation function. The experimentally derived and the predicted $\xi_p$-distributions for c-quark fragmentation are compatible with each other within the larger errors over the full considered region supporting the MLLA ansatz. There is no direct evidence for an excess multiplicity at large $\xi_p$ as seen in b-quark fragmentation but the errors are larger. 

The different behaviour of the c-quark and b-quark fragmentation functions is caused by the different behaviour of the $\xi_p$-distributions at the low energies $W_0(M_Q)$, i.e.\ at 2.6 and 8.0~GeV, respectively. In this way, the essential features of the dead cone effect are explained by the subtraction of particles with the $\xi_p$-distribution at the respective mass scale $M_Q$ from the full particle ensemble at energy $W$.

We also observe in Fig.~\ref{fig:ion as displayed in Fig. 1 (left panel) and the same exp_subtraction}, that the light and heavy quark spectra approach each other for large $\xi_p$ as theoretically expected, since the soft particles are mainly emitted at large angles and are thereby insensitive to the cut-off $\Theta_0=M_Q/E_Q$, i.e.\ the dead cone effect. 
 
It is noted, that our results on the shapes of observed uds-quark and MLLA-expected b-quark distributions in Fig.~\ref{fig:ion as displayed in Fig. 1 (left panel) and the same exp_subtraction} qualitatively agree with the expectations presented in the original publication~\cite{Dokshitzer:1991fc}.  
 
\begin{figure}[htpb!]
\begin{tabular}{cc}
\hskip -1.1cm
\includegraphics[height=8.5cm,width=9.5cm]{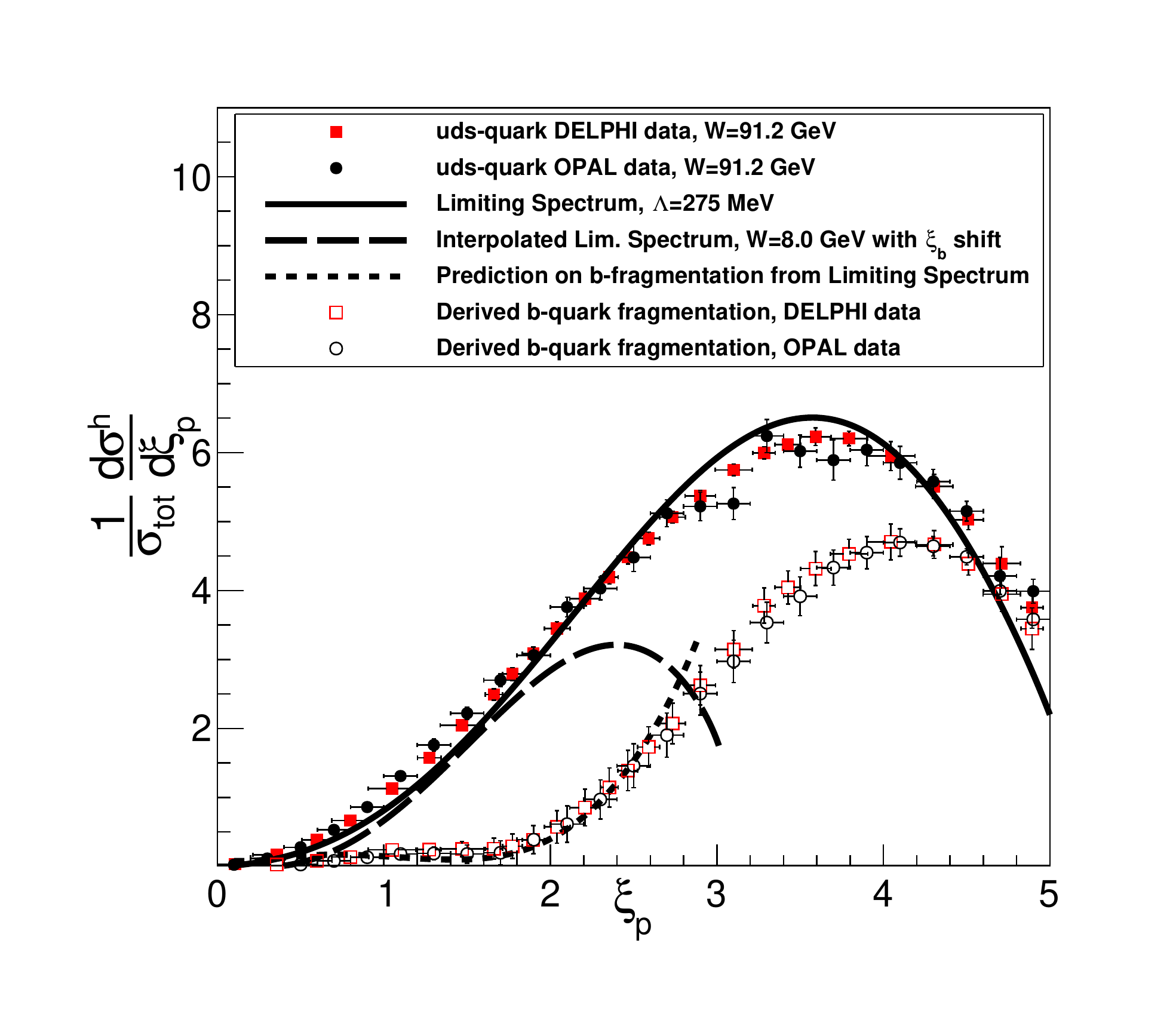} &
\hskip -1.0cm
\includegraphics[height=8.5cm,width=9.5cm]{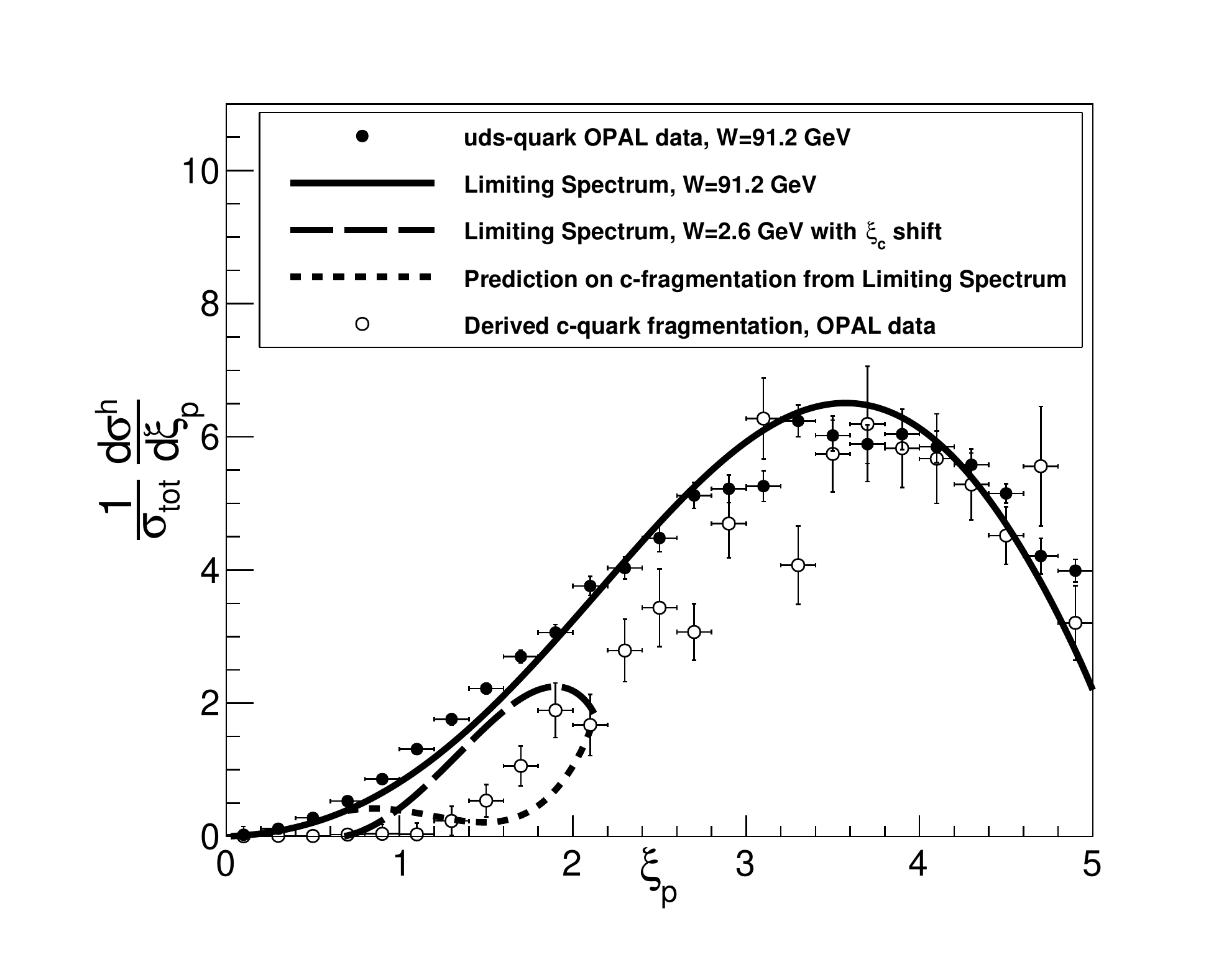} \\
\hskip -1.1cm
\includegraphics[height=8.5cm,width=9.5cm]{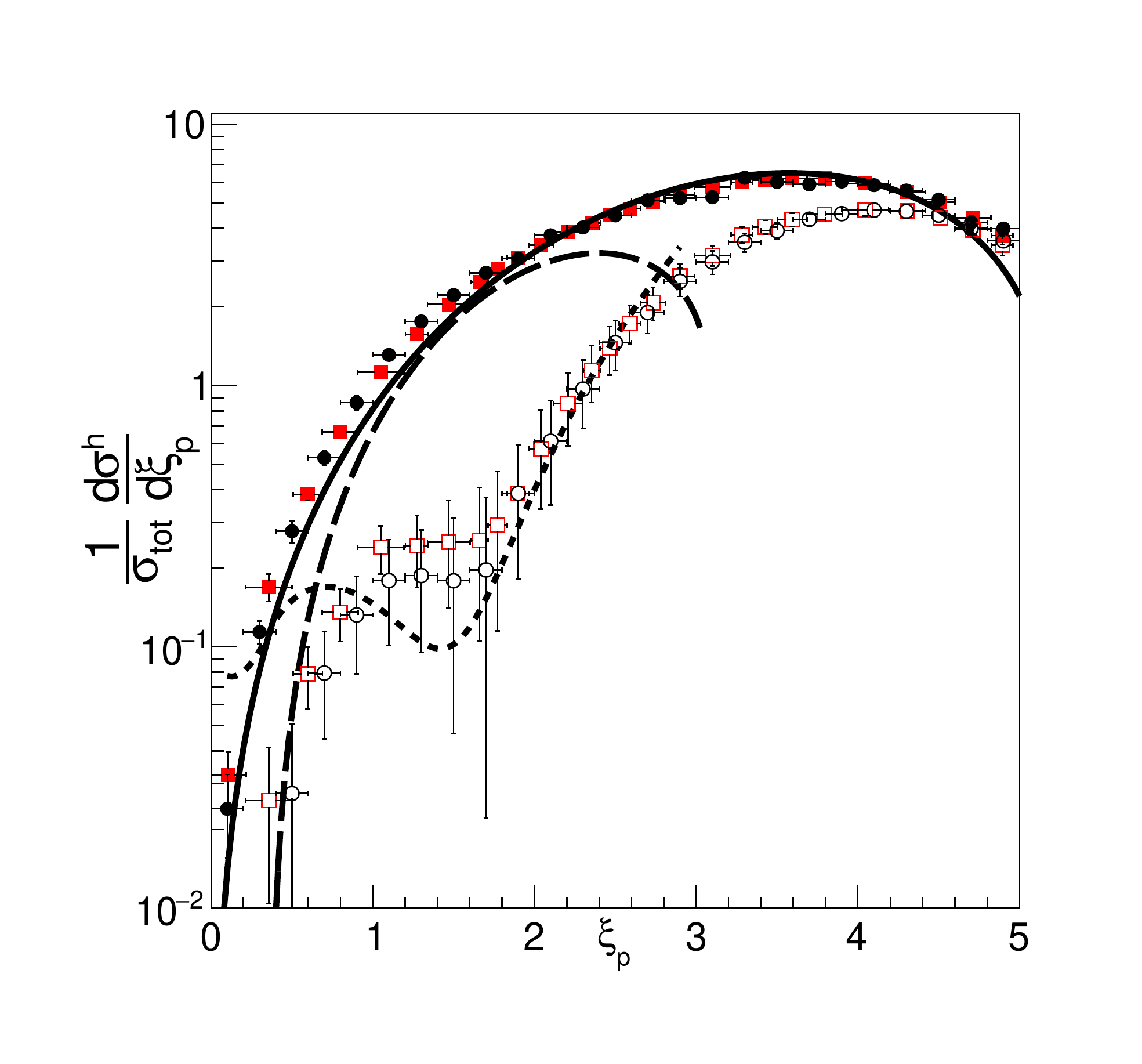} &
\hskip -1.0cm
\includegraphics[height=8.5cm,width=9.5cm]{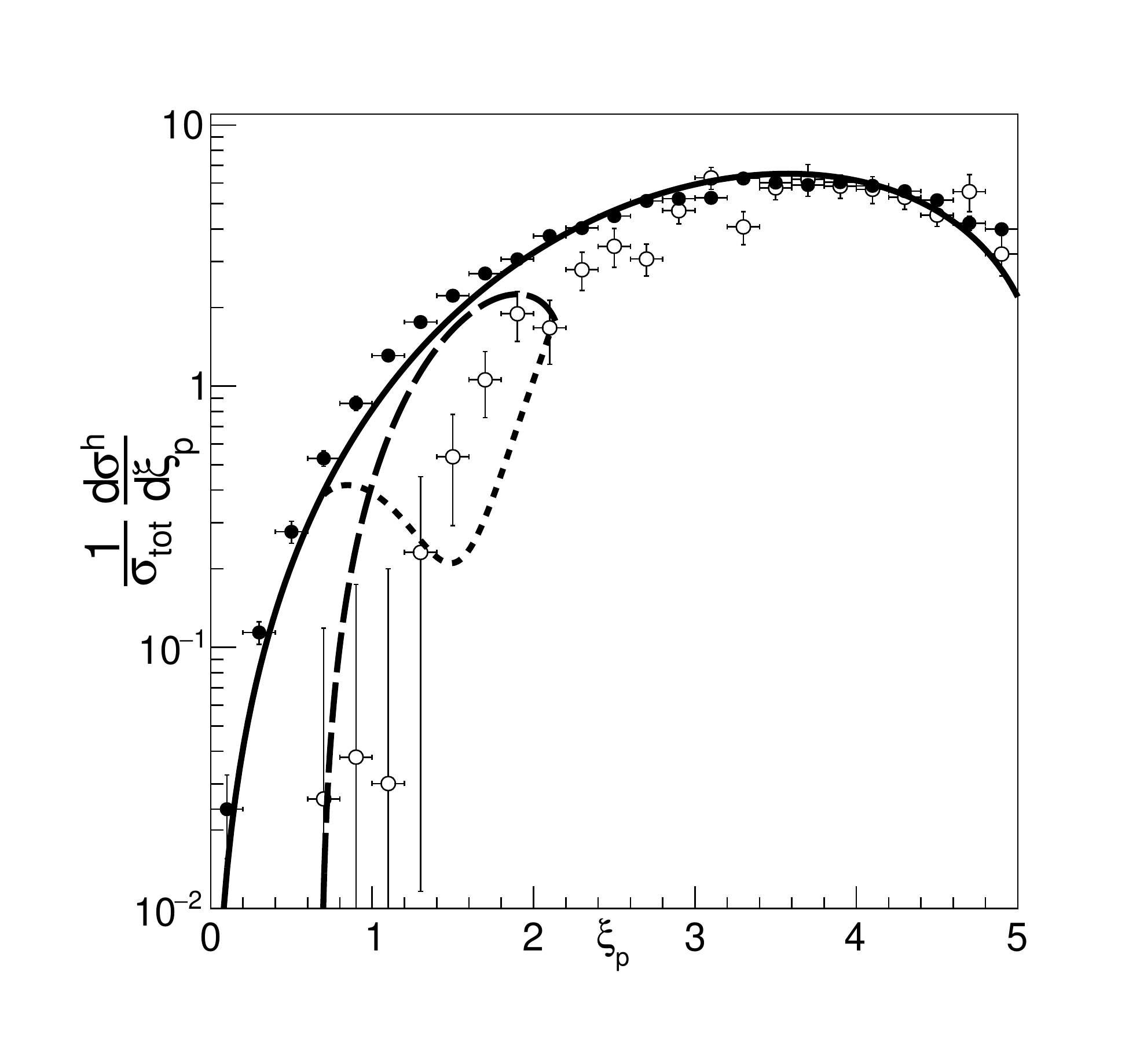} \\
\end{tabular}
\caption{Fragmentation function in $\xi_p=\ln(1/x_p)$ for uds-quarks at 91.2~GeV together with the MLLA Limiting Spectrum distributions; at $W_0=8.0$~GeV the Limiting Spectrum at this energy including the correction for charm decays (normalisation $K_{ch}=1.43$, App.~\ref{sec:interpolation}) and with shift $\xi_b=0.36$, see eq.~\eqref{xicbvalues}. Subtracting these distributions as in eq.~\eqref{MLLAeqxi} yields the MLLA prediction for the b-quark fragmentation function to be compared with the experimentally derived one (left panels); the corresponding results for c-quark fragmentation (right panels).}
\label{fig:th_subtraction}
\end{figure}

\subsection{Description of $\xi_p$-distributions by the Limiting Spectrum}
\label{sec:limitingspectrum}

In this subsection we will probe the MLLA suggested prediction eq.~\eqref{MLLAeqxi} by inserting, for the description of inclusive $\xi_p$-distributions, the analytical MLLA results for the $\xi$-distribution of gluons. In a particular approach, the transverse momentum cutoff is taken at its minimum value $Q_0=\Lambda$ and one obtains the so-called ``Limiting Spectrum'', which can be written in terms of an integral representation of the confluent hypergeometric function~\cite{DKMT-book}. In this application of the MLLA framework, one assumes that the parton cascade evolves down to the hadron mass scale $Q_0=\Lambda$ and represents the hadron cascade in an average sense (``Local Parton Hadron Duality'' (LPHD)~\cite{Azimov:1984np}). As the bulk of particles inside the Gaussian hump have only small momenta of a few GeV, the number of active flavours in the calculation of the coupling $\alpha_s$ is usually taken as $n_f=3$~\cite{DKMT-book,OPAL:1990vmr,Albino:2004xa}. The only remaining parameters in this approach are the QCD scale $\Lambda$ and an overall normalisation factor $K_{ch}$. Previous fits to the charged particle spectra for all flavours yielded values, see e.g.~\cite{OPAL:1990vmr}, with $\Lambda=250$~MeV and $K_{ch}=1.28$ at c.m.s.\ energy $W=91.2$~GeV. Distributions at lower energies down to 14~GeV could be fitted with the same $\Lambda$ but for $K_{ch}$ increased by 14$\%$. This energy dependence is interpreted as an effect from higher order corrections to the MLLA.

The Limiting Spectrum function will now be used for the description of the $\xi_p$ spectra in this analysis. The $\xi_p$-distribution for b-quark fragmentation at 8~GeV is obtained by using the same parameters as for the neighbouring energies and applying a correction for charm quark production (see App.~\ref{sec:interpolation}). This spectrum at 8~GeV is shown in 
Fig.~\ref{fig:th_subtraction} (left panel) as the dashed line, also shown is the corresponding fit to the uds-data at $W=91.2$~GeV. The parameters of the fit are obtained as $\Lambda=275$~MeV in this energy range, $K_{ch}=1.28$ at $W=91.2$~GeV and $K_{ch}=1.52$ at the lower energies. From these two Limiting Spectrum distributions one obtains the MLLA based predictions for the b-quark fragmentation function using eq.~\eqref{MLLAeqxi} shown as the short dashed line. This prediction is in quantitative agreement with the experimentally derived b-quark spectrum (see Sec.~\ref{sec:Reconstruction}) in the region around the maximum of the spectrum at 8~GeV for $\xi_p \gtrsim 1.8$, with a mild disagreement below that value where errors are large and the relation eq.~\eqref{MLLAeqxi} is only approximately valid (at large x). The predictions end at the kinematic limit of the Limiting Spectrum $\xi_0=\xi_{max}^{lim}+ \xi_b$ = 3.05 with $\xi_{max}^{lim}=\rm{ln}(W/(2Q_0))$. 

The same procedure is followed with the c-quark fragmentation in the right panel of Fig.~\ref{fig:th_subtraction}. At the low energy of 2.6~GeV both parameters have to be adjusted and are reported by BES~\cite{BES:2003xdf} as $\Lambda=342$~MeV and $K_{ch}=1.52$. In the narrow available region in $\xi_p$ the predictions at the energies 2.6 and 91.2~GeV are nearby but somewhat below the data.

\begin{figure}[tpb!]
\begin{tabular}{cc}
\hskip -1.1cm
\includegraphics[height=9.0cm,width=9.5cm]{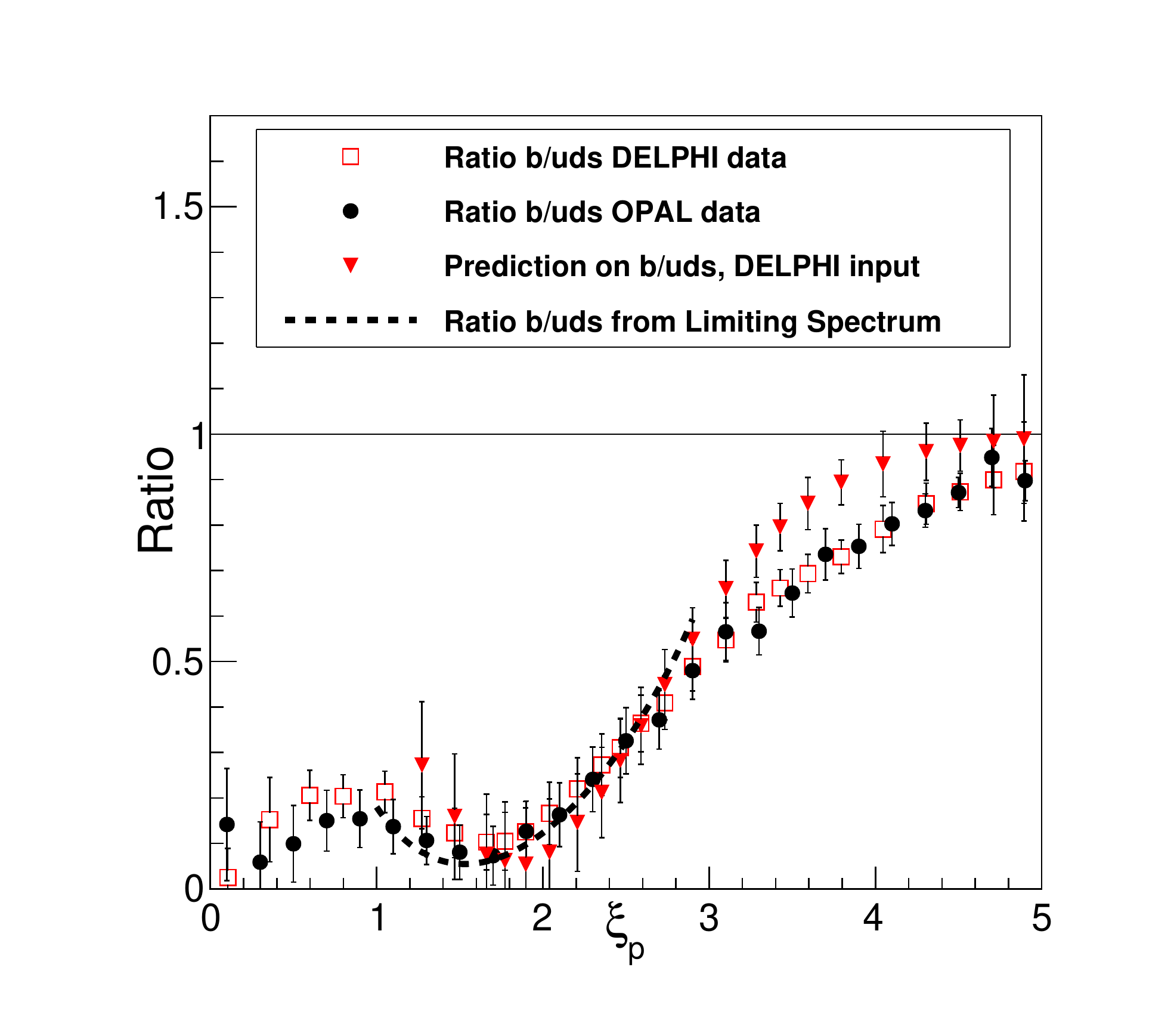} &
\hskip -1.0cm
\includegraphics[height=9.0cm,width=9.5cm]{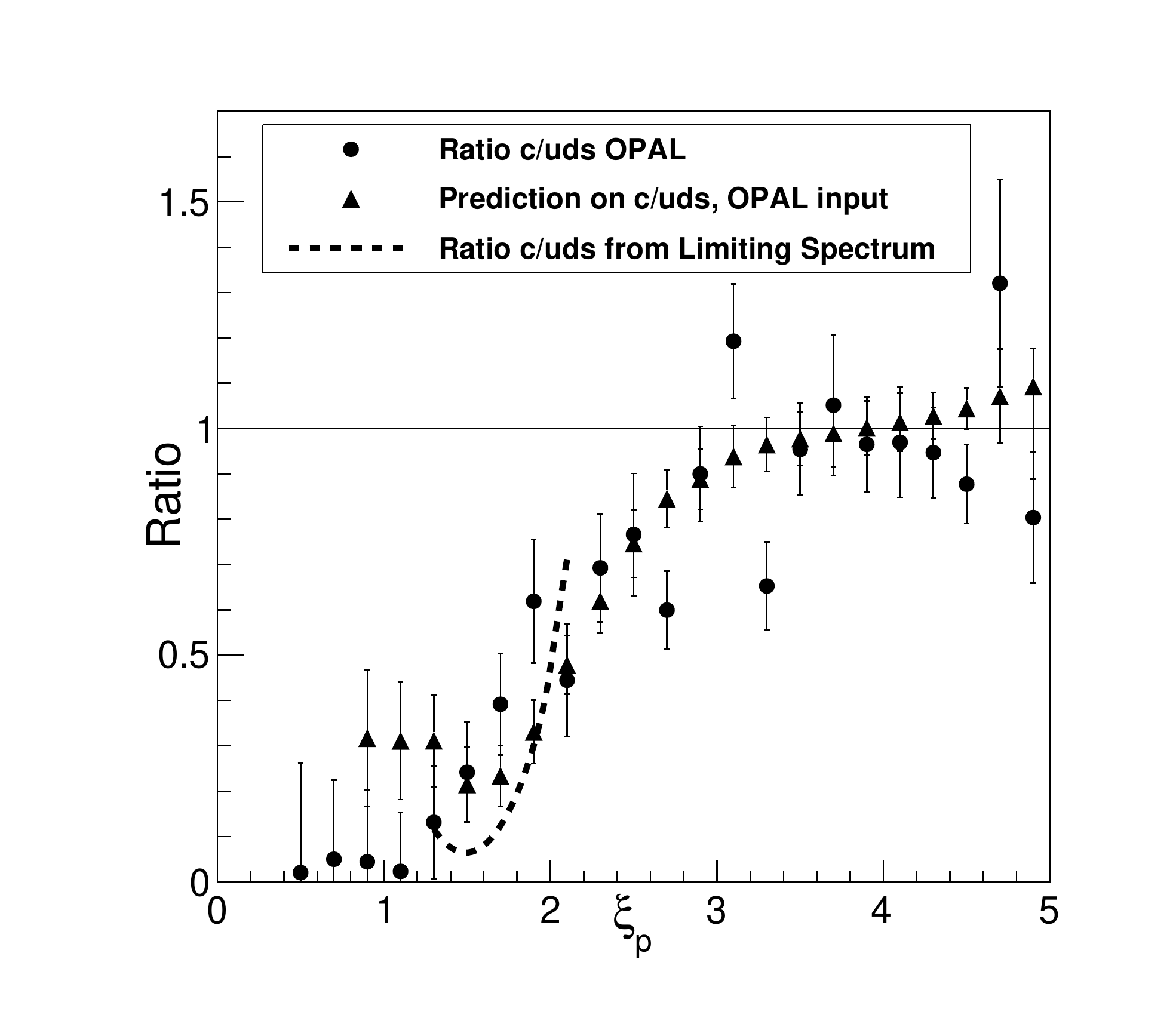} \\
\end{tabular}
\caption{Ratio of the heavy b-quark over the light uds-quark fragmentation functions (left panel) and the corresponding ratio for c-quark (right panel) together with the MLLA expectations based on comparison with experimental data and with Limiting Spectrum distributions.}
\label{fig:exp-deadcone_ratios}
\end{figure}

In Fig.~\ref{fig:exp-deadcone_ratios} we show again the ratio of heavy quark over light quark fragmentation functions in comparison with the MLLA prediction eq.~\eqref{MLLAeqxi} using data input and Limiting Spectrum fits. These ratios are correctly reproduced in the central region $1\lesssim\xi_p\lesssim 3$ for b-quarks and in $1\lesssim\xi_p\lesssim 2$ for c-quarks. 
Below $\xi_p \sim 1$ ($x\gtrsim 0.4$) the Limiting Spectrum ratios are rising again, because of a mismatch in the lower limit in $\xi_p$ for the limiting spectrum at 91.2~GeV and the shifted one at the low energy $W_0$ (i.e.\ at 2.6 or 8~GeV), therefore we excluded those results from the figure (see also next section). The amount of suppression from the dead cone effect is correctly reproduced for both heavy quark fragmentation processes.
 
The figures~\ref{fig:th_subtraction}, \ref{fig:exp-deadcone_ratios} and \ref{fig:theoretical_curve} show the good overall description of the $\xi_p$-distributions for both light and heavy quarks within the very compact MLLA-LPHD and Limiting Spectrum ($Q_0=\Lambda$) approach in terms of only two parameters, the QCD scale $\Lambda$ and the slowly moving normalisation parameter $K_{ch}$. For the very low energy around 2~GeV also a change of $\Lambda$ from 275 to 340~MeV is required. These small variations reflect the relevance of higher order corrections beyond MLLA. 

\subsection{Behaviour of fragmentation functions near kinematic limits}
\label{sec:convolution}

In the figures~\ref{fig:ion as displayed in Fig. 1 (left panel) and the same exp_subtraction} and~\ref{fig:th_subtraction}, it is demonstrated that the MLLA expectation eq.~\eqref{MLLAeqxi} quantitatively predicts the suppression of particle production in the central region around the maximum of the $\xi_p$-distribution at the lower mass scale $W_0$. For the b-quark, however, there is a major surplus of particles at large $\xi_p$ beyond expectation and a smaller excess at small $\xi_p$. To quantify these effects more clearly, we investigate the difference between the heavy and light quark fragmentation functions at $W=91.2$~GeV which, according to the MLLA expectation eq.~\eqref{MLLAeqxi}, should just yield the expected light quark $\xi_p$-distribution at the lower mass scale $W_0$. This will clarify how the predicted $\xi_p$- distribution at the low energy $W_0$ deviates from the observed one. 

\begin{figure}[tpb!]
\begin{tabular}{cc}
\hskip -1.1cm
\includegraphics[height=10.5cm,width=9.5cm]{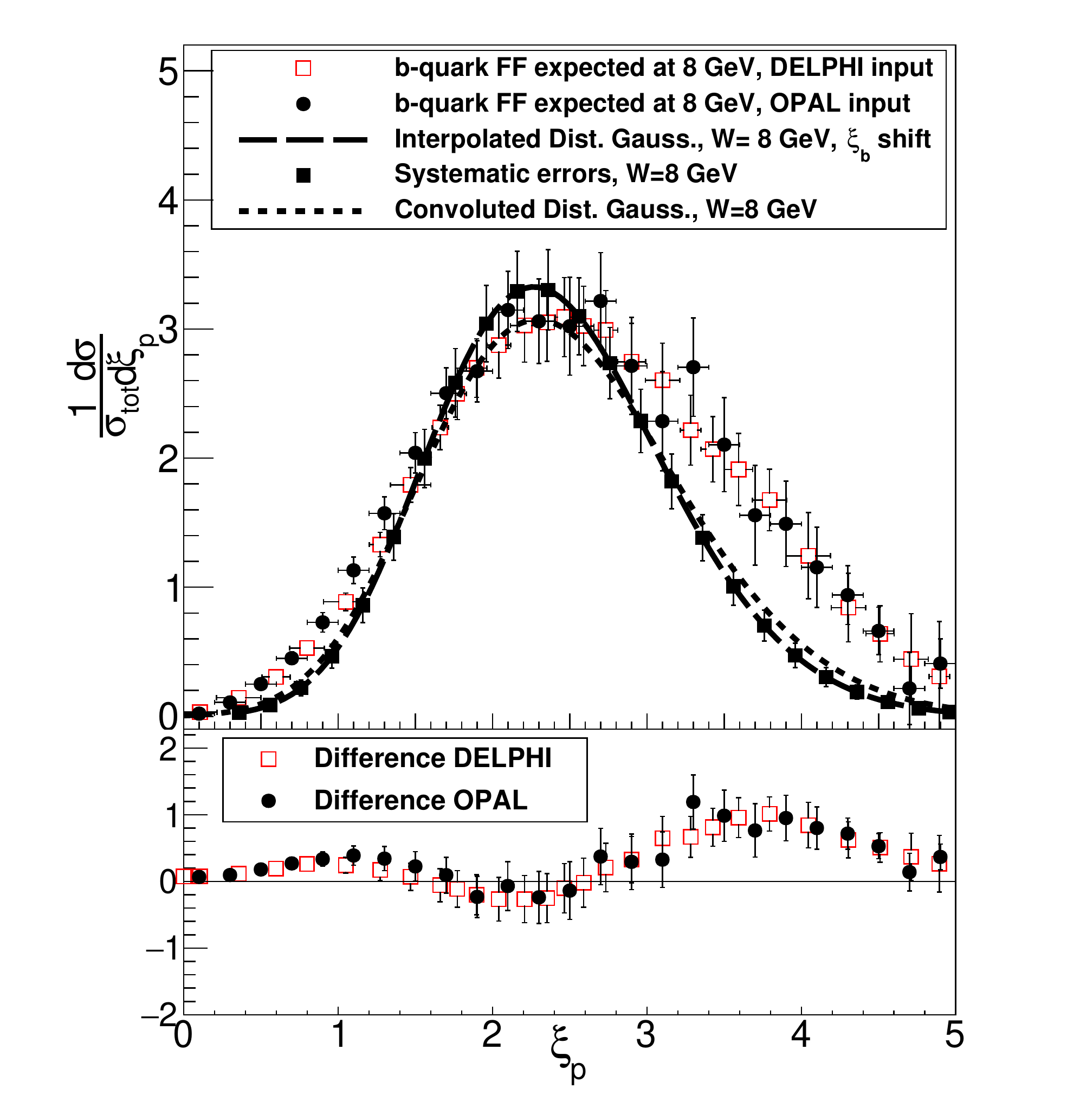} &
\hskip -1.0cm
\includegraphics[height=10.5cm,width=9.5cm]{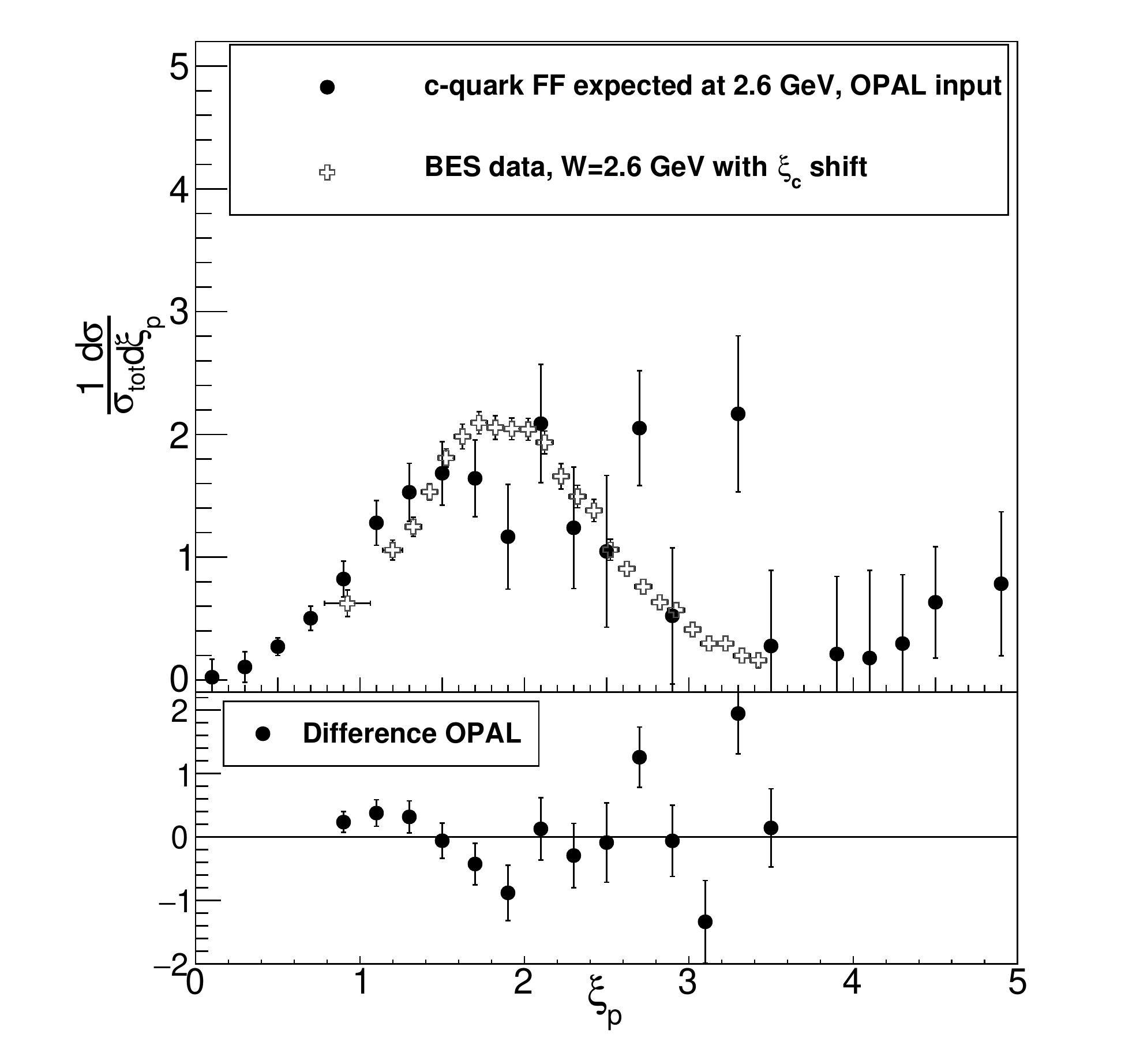} \\
\end{tabular}
\caption{Expected $\xi_p$-Fragmentation Function (FF) at $W_0=8.0$~GeV constructed according to MLLA eq.~\eqref{MLLAeqxi} as difference of $\xi_p$-distributions for uds-quark and b-quark jets with DELPHI and OPAL data as input; compared in absolute normalisation with interpolated Distorted Gaussian (DG) at 8~GeV with shift $\xi_b=0.36$ and the same DG distribution convoluted with the leading particle $x_b$-spectrum of the heavy b-quark (left panel), the corresponding results for the expected $\xi_p$-Fragmentation Function at $W_0=2.7$~GeV from uds-quark and c-quark jets in comparison with BES data (right panel).}
\label{fig:convolution}
\end{figure}

These differences are shown in Fig.~\ref{fig:convolution} for the b-quark (left panel) and c-quark fragmentation (right panel). The MLLA expected $\xi_p$-distributions are compared with the experimental $\xi_p$-distribution: for the b-quark at 8~GeV with the distribution obtained by interpolation with corresponding systematic errors (dashed line, see sec.~\ref{sec:distortedgaussion}), for the c-quark with the observed distribution at 2.6~GeV by the BES collaboration. The differences between the MLLA expected and experimental $\xi_p$-distributions are shown in the lower part of Fig.~\ref{fig:convolution}. 

This figure clearly shows that the difference between heavy and light quark fragmentation, in its main features, can just be related to the low energy ``hump backed plateau'' which changes with the MLLA mass scale $W_0= \sqrt{e} M_Q$. When the energy $W_0$ is increased from 2.6 to 8.0~GeV, the $\xi_p$-distribution shifts to a higher mean value $\bar \xi_p$ with larger width and increasing height in agreement with the behaviour known from experiment in absolute terms. This result not only explains the limits of their ratios in Fig.~\ref{fig:exp-deadcone_ratios} for small and large $\xi_p$ with $R\sim 0$ and $R=1$, but also the behaviour in between. 

For the b-quark, the interpolated distribution at 8~GeV approaches quite closely the data in the central region $1\lesssim \xi_p \lesssim 3$, but falls somewhat below the expectations for the very small $\xi_p\lesssim 1$ ($x_p\gtrsim 0.4$) and there is a considerable and very significant excess over the expectation in the large $\xi_p\gtrsim 3$ region. For the c-quark fragmentation there is a good agreement but errors become large for the larger $\xi_p$. 

\begin{figure}[htpb!]
\begin{tabular}{cc}
\hskip -1.1cm
\includegraphics[height=10.5cm,width=9.5cm]{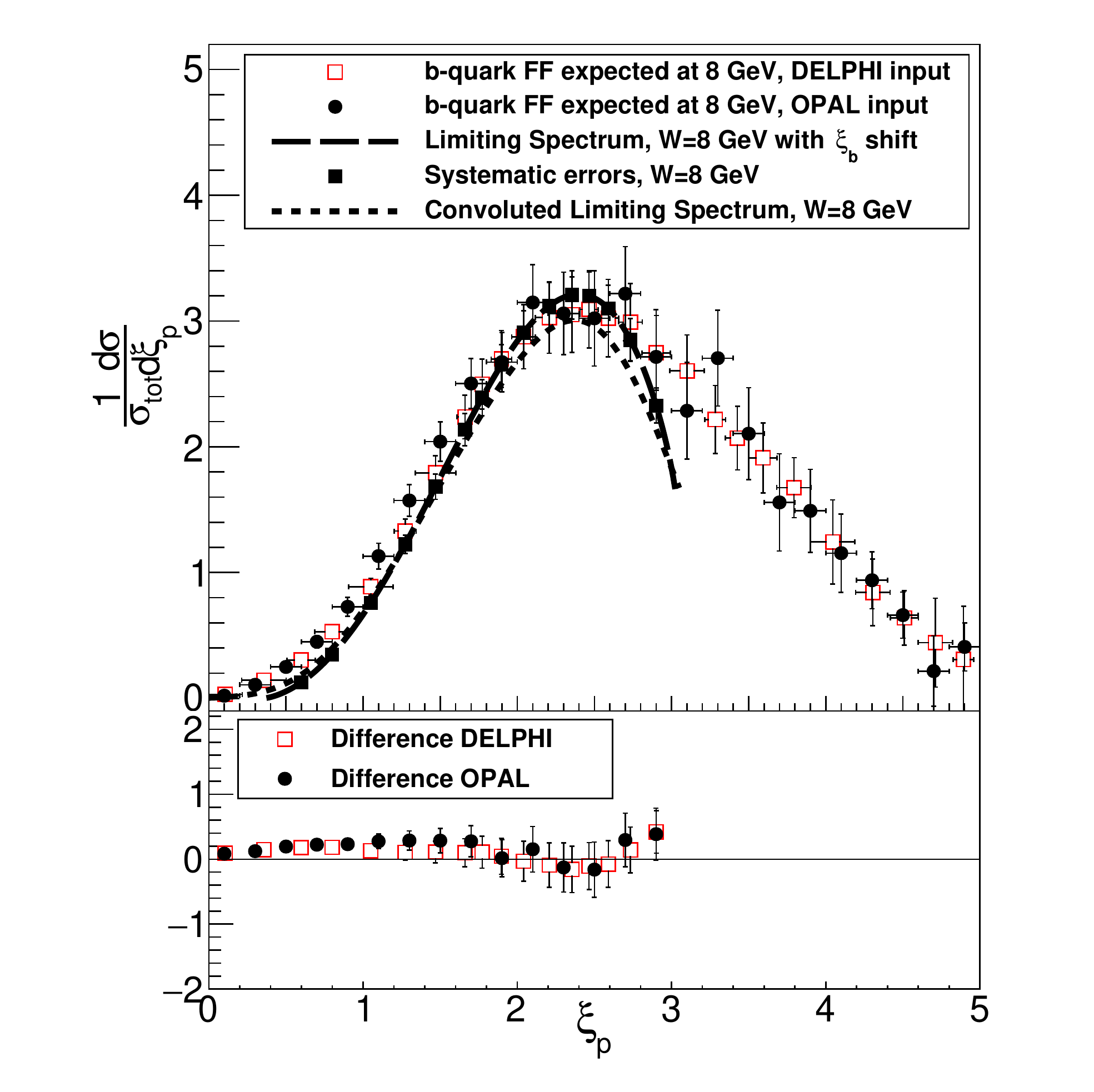} &
\hskip -1.0cm
\includegraphics[height=10.5cm,width=9.5cm]{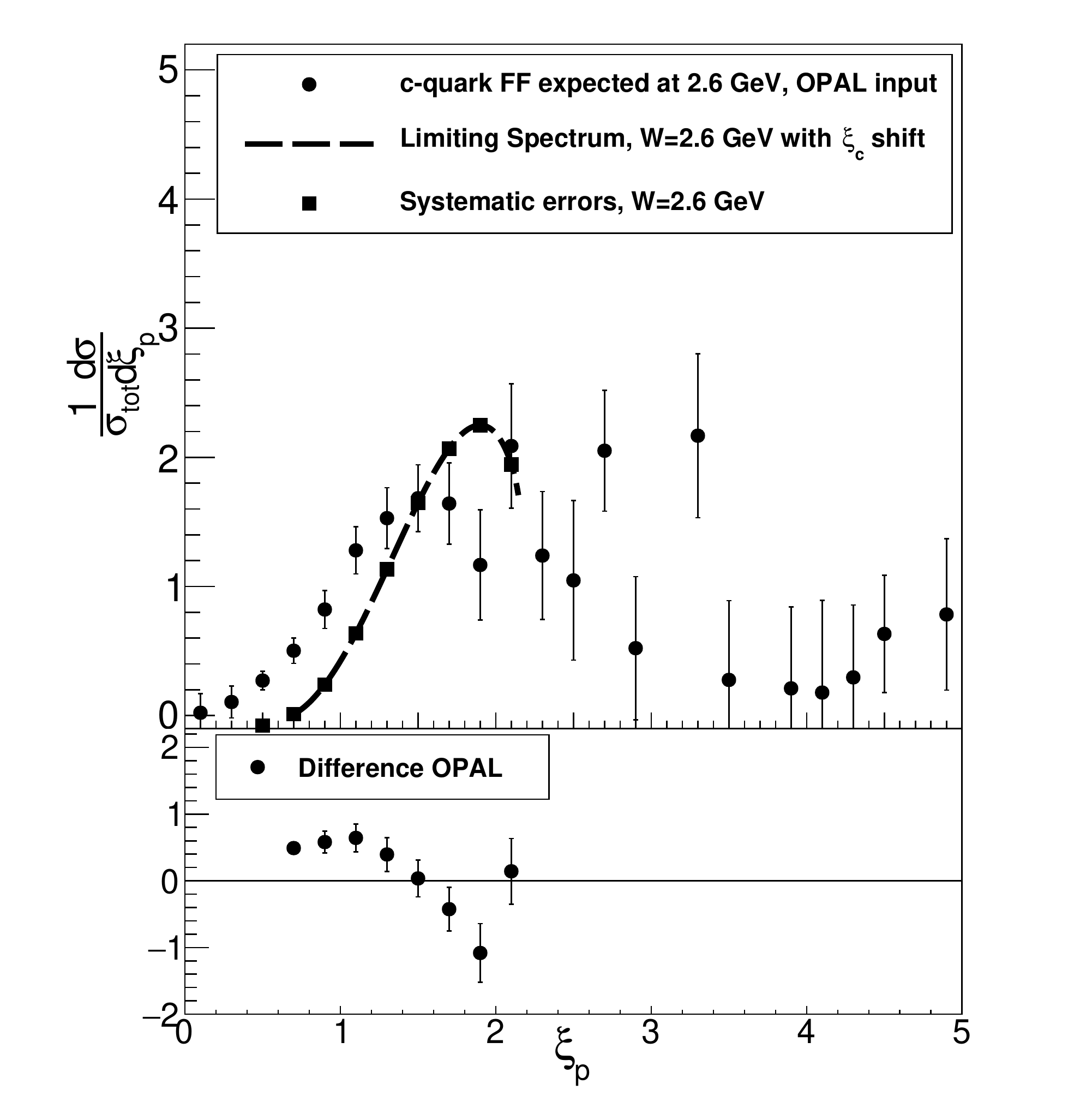} \\
\end{tabular}
\caption{Expected $\xi_p$-Fragmentation Function (FF) at $W_0=8.0$~GeV constructed according to MLLA eq.~\eqref{MLLAeqxi} as difference of $\xi_p$-Fragmentation Functions for uds-quarks and b-quarks; compared in absolute normalisation with interpolated Limiting Spectrum at 8.0~GeV with shift $\xi_b=0.36$, see eq.~\eqref{xicbvalues} and the same Limiting Spectrum convoluted with the leading particle $x_b$-spectrum of the heavy b-quark (left panel); the corresponding results for c-quarks with Limiting Spectrum calculations using parameters by BES and $\xi_c=0.6$, see eq.~\eqref{xicbvalues} (right panel).}
\label{fig:convolution_1}
\end{figure}

We also compare these experimental results with the MLLA Limiting Spectrum distributions in Fig.~\ref{fig:convolution_1}.
The agreement for b-quarks is rather satisfactory up to the region close to the upper limit at $p\sim \Lambda$ i.e.\ at $\xi\sim 3$, but there are deviations at the very small $\xi_p\lesssim 1$ ($x_p \gtrsim 0.4$). For c-quarks the agreement in the central region around $\xi_p\sim 2$ is satisfactory but there are considerable differences at small $\xi_p$. These deviations at small $\xi_p$ are a consequence of the approximate form of the eq.~\eqref{MLLAeqxi} in which the $\xi_p$-distribution is shifted to $\xi\gtrsim \xi_Q$, i.e.\ to the values 0.36 and 0.70 for b- and c-quarks, respectively. Therefore the approximate form of eq.~\eqref{MLLAeqxi} works best in the central region around the maximum, away from limits $\xi_p=0$ and $\xi_p=\xi_{\rm max}^{\rm lim}=\ln(W/(2\Lambda))$.

It appears that the width of the observed distribution in b-quark fragmentation is larger than the expected one. One possible explanation could be that the momentum fluctuations of the heavy quark are larger than anticipated in the MLLA formula eq.~\eqref{MLLAeq} where a fixed energy loss $\langle x_Q\rangle$ is assumed. More generally, one could consider the emitting heavy quark with a distribution $F_Q(x_Q)$ or 
\begin{equation}
  \bar F_Q(\xi_Q)= x_Q F_Q(x_Q),\ \ \xi_Q = \rm{ln}(1/x_Q).
\label{FQdef}
\end{equation}
Then the distribution of the final partons is obtained from the convolution integral
\begin{equation}
  \bar D_q(\xi_p,W_0)= \int_0^{\xi_p}\ d\xi_p' \bar F_Q(\xi_p') \bar D_q(\xi_p-\xi_p', W_0), \ \ W_0=8 \ \text{GeV},
\label{convol}
\end{equation}
which may also fill the region $\xi_p<\xi_Q$. With $F_b(x_b)=\delta(x_b-\langle x_b\rangle)$, the eq.~\eqref{MLLAeq} is restored. If one takes, as an exercise, for the momentum spectrum of the heavy quark $F_b(x_b)$, the distribution as experimentally measured (DELPHI~\cite{DELPHI:2011aa}), the second curve (short dashed) in Fig.~\ref{fig:convolution} and Fig.~\ref{fig:convolution_1} from the convolution is obtained, which is broader than the experimental $\xi_p$-distribution for 8~GeV, as expected, but the effect is rather small and cannot explain the observed deviations. 

Finally, the MLLA prediction for b-quark fragmentation and the experimental spectrum at 8~GeV are compared separately for the regions below and above $\xi_p = 3$. To this end, the difference between the expected and experimental distributions at 8~GeV, see Fig.~\ref{fig:convolution}, is fitted to a polynomial function, and the respective multiplicities $\Delta N=N(8\ \rm{GeV})^{MLLA}$ - N(8~GeV)$^{\rm exp}$ are calculated from the integrals over the two $\xi_p$ regions for Distorted Gaussian (Fig.~\ref{fig:convolution}) and the Limiting Spectrum distributions  (Fig.~\ref{fig:convolution_1}). The results are shown in Tab.~\ref{tab:parameters}.

\begin{table}[!bhtp]
\caption{Integrals $\Delta N$ over the difference data shown in Fig.~\ref{fig:convolution} and Fig.~\ref{fig:convolution_1} (left panels) between expected and experimental $\xi_p$-distributions for b-quark fragmentation for different $\xi_p$ regions and two interpolating functions at 8~GeV.}
\label{tab:parameters}
\setlength{\tabcolsep}{2.5pc}
\begin{ruledtabular}
\begin{tabular}{ccc}
$\xi_p$-range &  Distorted Gaussian &  Limiting Spectrum \\
\colrule
$\rm{all} \ \xi_p$ & $\Delta N=1.52\pm0.25$  & $\Delta N=0.24\pm0.15$   \\
$\xi_p<3$ & $\Delta N_{\rm low}=0.14\pm0.16$  & $\Delta N_{\rm low}=0.24\pm0.15$   \\
$\xi_p>3$ & $\Delta N_{\rm high}=1.37\pm0.16$  & $-$ \\
\end{tabular}
\end{ruledtabular}
\end{table}

The first number $\Delta N$ in Tab.~\ref{tab:parameters} is obtained by summing the data points over the full $\xi_p$ range and it should agree with the results using the published total multiplicity data instead. For DELPHI, one has N(8~GeV)$^{\rm MLLA}$ = $N_{uds}-N_b+n_{\rm dec} = (19.44\pm 0.34)-(23.17\pm 0.38)+(11.10\pm0.18)=7.87\pm0.54$ and with $N_{uds}(8\ \rm {GeV})^{exp}$ = $6.1\pm0.3$, one finds $\Delta$N =$1.8\pm 0.6$, which compares well with the value from our fit $\Delta$N$=1.52\pm 0.25$. 

As a main result for the b-quark fragmentation, it can be seen from Tab.~\ref{tab:parameters} that the full multiplicity in the lower part of the $\xi_p$-spectrum (its integral over $\xi_p<3$) agrees with the MLLA expectation for both interpolating functions, i.e.\ $\Delta N=0$, but at large $\xi_p$ there is a significant difference for which we have no direct explanation, but the effect appears in the kinematic region $p \lesssim Q_0=\Lambda$ outside the validity of the perturbative MLLA approach. 

In this way, we also suggest a solution to the problem found in the previous study of full multiplicities~\cite{Dokshitzer:2005ri}, where a moderate but significant discrepancy between expected MLLA results and the experimental finding was noted, namely $\delta_{b\ell}^{\rm MLLA}-\delta_{b\ell}^{\rm exp}= 1.26\pm 0.42$ multiplicity units, see eq.~\eqref{delta}, corresponding to our result on the equivalent quantity $\Delta N= 1.52\pm 0.25$. The observed difference of these two numbers comes from the slightly different determination of N(8~GeV). We now conclude that the previously observed discrepancy in the MLLA multiplicity equation~\eqref{dc-multiplicity} can be related to the contribution from the ultrasoft particles with $\xi_p>3$. These are contributions from other sources outside the control of perturbation theory for the heavy quark jets.

For the c-quark fragmentation the statistical errors in figures~\ref{fig:ion as displayed in Fig. 1 (left panel) and the same exp_subtraction} and~\ref{fig:convolution} are too large to confirm or exclude such an ultrasoft anomaly. The result on multiplicities 
$\delta_{c\ell}^{\rm MLLA}-\delta_{c\ell}^{\rm exp}= 0.5\pm 0.6$ from Eq.~\eqref{deltac} does not show any such effect either.

\subsection{Sensitivity of experimental dead cone data to the heavy quark mass}
\label{sec:heavy_quark_mass}

In the last subsection we have compared the experimental data on the difference between light and heavy quark fragmentation in figures~\ref{fig:convolution} and~\ref{fig:convolution_1} with the expectation for the fragmentation function at the given low energy scale $W_0=\sqrt{e}M_Q$ according to the MLLA estimate eq.~\eqref{MLLAeqxi}. In turn, one could treat the low energy scale $W_0$ in this equation as a free parameter to be determined from the best fit of the $\xi_p$ spectrum in $e^+e^-$ annihilation to the experimental data in Fig.~\ref{fig:convolution}. We perform such a fit to the data in the lower pad of Fig.~\ref{fig:convolution} and determine a constant shift to the zero line (corresponding to the $\xi_p$-distribution at $W_0=8.0$~GeV) in the central region $1.6\lesssim \xi_p\lesssim 2.6$ avoiding contributions from the excess multiplicity at large~$\xi$. The maximum height of the inclusive $\xi_p$ spectra $h_{\rm max}(W)$ as measured by BES~\cite{BES:2003xdf} and TASSO~\cite{TASSO:1990cdg} collaborations is found by interpolation to rise by 0.2 units for an increase of $W$ by 1~GeV near $W=8.0$~GeV. For the DELPHI and OPAL data we obtain the shifts $\delta h_{\rm max}=-0.17 \pm 0.13$ and $\delta h_{\rm max}=-0.14 \pm 0.15$, respectively, or combined $\delta h_{\rm max}=-0.16\pm 0.10$ and correspondingly for the low energy scale  
\begin{equation}
  W_0^{\rm exp}=(7.2\pm 0.5) \ \rm{GeV}\ \rm{ (DELPHI \  and \ OPAL)}.
\label{w0_determination}
\end{equation}

This is to be compared with the evaluation~\cite{Dokshitzer:2005ri} $W_0^{\rm MLLA}=(8.0 \pm 0.2)$~GeV, based on the b-quark pole-mass $M_b=4.85 \pm 0.15$~GeV and the MLLA correction $\sqrt{e}=1.65$ applied to the lowest order DLA result $W_0^{\rm DLA}=M_Q$. 
It is remarkable, how close the MLLA prediction with its large correction factor comes to the experimental data. If we would take the quark mass $M_b$ itself as low energy scale we had to replace the Gaussian curve for $W_0=8.0$~GeV in Fig.~\ref{fig:convolution} by the curve at 4.8~GeV, displayed in Fig.~\ref{fig:theoretical_curve}, which is clearly far away from the data in Fig.~\ref{fig:convolution}. The experimental uncertainty of the low energy scale $W_0$ in~\eqref{w0_determination} is about 7$\%$, but there are also theoretical uncertainties from possible corrections beyond MLLA to the correction factor for the energy scale $\sqrt{e}$ and from the approximation of applying a constant shift in eq.~\eqref{MLLAeqxi}. 
Therefore, although the dead cone measurements at present cannot be competitive to the available heavy quark mass determinations, they come close to the predicted value within less than 10$\%$. 

\section{Conclusions}
\label{sec:conclusions}

The dead cone effect predicted by perturbative QCD has been studied using data taken at LEP on identified heavy b- and c-quark and light uds-quark fragmentation. 
The dead cone effect for particle production at small angles to the primary quark is also reflected in the production of large momenta, as is typical for the jet structure. In the present study, QCD expectations for the momentum spectra in heavy quark jets based on the MLLA~\cite{Dokshitzer:1991fc,Dokshitzer:1991fd} are investigated. 

At first, we reconstruct the inclusive distributions of charged particle momenta using the variable $\xi_p=\ln(1/x_p)$ in b-quark and c-quark events by correcting for B-hadron and Charm-hadron decays. In the comparison of heavy and light quark fragmentation we observe a convergence of the spectra for large $\xi_p$ ($x_p\to 0$) but a strong suppression of the fragmentation functions of the heavy b- and c-quarks with respect to the one of the light uds-quarks with decreasing $\xi_p\to 0$ (increasing $x\to 1$) down to a fraction of $\lesssim 1/10$. This observed almost complete suppression reflects the presence of the dead cone with a high significance ($\gg 5\sigma$). There is a characteristic difference between b- and c-quark fragmentation, in that the decrease for the c-quark is shifted towards lower $\xi_p$ as compared to the b-quark. It would be desirable to replace the MCEG based subtraction of the charged heavy hadron decay products by an experimental measurement in order to remove any residual model dependence.

The $\xi_p$-distributions derived from experimental data are then compared directly with the QCD expectations within the MLLA following the hypothesis of Local Parton Hadron Duality (LPHD). This QCD analysis provides a quantitative explanation of the dead cone effect: the difference between the heavy and light quark fragmentation functions in the variable $\xi_p$ at high c.m.s.\ energy $W$ is just given by the $\xi_p$-fragmentation function at the lower energy $W_0=\sqrt{e}M_Q$ with the heavy quark mass $M_Q$ (see eq.~\eqref{MLLAeq}).

The equation~\ref{MLLAeq}, an estimate within MLLA, is tested first with the experimentally observed or derived $\xi_p$-distributions as input. For both the b-quark and the c-quark fragmentation this equation is found to be well supported in the central kinematic region around the peak of the $\xi_p$-spectrum at scale $M_Q$ corresponding to the momentum range $x\lesssim 0.4$ and $p\gtrsim \Lambda$. It explains quantitatively the suppression of both fragmentation functions down to about $1/10$ of the one for uds-quarks. The different suppression profiles of c- and b-quark fragmentation are directly related to the different shapes of the $\xi_p$ spectra (the ``hump-backed plateau'') at the respective c- and b-quark mass scales $W_0$, i.e.\ at the c.m.s.\ energies 2.7 and 8.0~GeV respectively.
In the MLLA estimate eq.~\eqref{MLLAeq}, the mean fractional momentum $\langle x_Q \rangle$ of the heavy quark appears as additional (known) parameter. This parameter is important for the successful quantitative description. The interplay between heavy quark fragmentation and energy loss deserves further attention.

The MLLA estimate eq.~\eqref{MLLAeq} has also been tested using as input the analytic expressions for the $\xi$-distributions obtained within MLLA with the simplification $Q_0=\Lambda$ for the $p_T$-cut off and the QCD scale, the so called ``Limiting Spectrum''. This formula describes to a reasonable approximation the experimental $\xi_p$-spectra in the c.m.s.\ energy range 2-100~GeV for allowed momenta $p>Q_0=\Lambda$ in terms of only two parameters, the QCD scale $\Lambda$ and the normalisation $K_{ch}$. These parameters show a small variation with energy which hints towards contributions beyond MLLA. In this way, a very compact representation of momentum spectra with two parameters for light and heavy quark jets in a wide kinematic region for not too large x ($x\lesssim 0.4$) and $p>\Lambda$ is established.

In the kinematic region of small $\xi_p\lesssim 1$ ($x_p\gtrsim 0.4$), at low values for the heavy quark fragmentation functions, violations of the MLLA relation have been observed. This is to be expected in the present approximate scheme using a shifted spectrum at the low energy $W_0$. Integrating the fragmentation function of the b-quark
over the important region $\xi_p\lesssim 3$ ($p\gtrsim \Lambda$) yields the respective multiplicity which is found in good agreement with the MLLA expectation. 
A large and significant excess of particle production in b-quark fragmentation over these expectations is observed for large $\xi_p\gtrsim 3$ which concerns the region of very soft particle production with $p\lesssim \Lambda$. 
The total excess multiplicity in this kinematic region corresponds quantitatively to the excess over MLLA expectations already noted in the previous study of the full multiplicity~\cite{Dokshitzer:2005ri}. This discrepancy comes from a kinematic region outside the validity of the perturbative approach. In case of the c-quark fragmentation 
no such excess at large $\xi_p$ can be resolved within the larger errors and there is a satisfactory agreement between prediction and experimental data for not too large momenta ($\xi_p\gtrsim 1$).

We estimated a value for the low energy scale $W_0=(7.2\pm0.5$)~GeV of the dead cone subtraction from the data which is consistent within the uncertainties with the MLLA prediction $W_0=\sqrt{e}M_b=(8.0\pm0.2)$~GeV. By relating $W_0$ to the b-quark mass this shows the mass sensitivity of the dead cone effect, which is, however, currently limited by uncertainties of the MLLA predictions.

We would like to point out that our results could be of direct relevance to the experimental task of identifying (tagging) jets originating from heavy quarks. Traditional heavy quark jet tagging algorithms only use variables derived from particles associated with the heavy hadron decay, see e.g.~\cite{Barker:2010pva} for a review of algorithms used at LEP. Recent developments using advanced machine learning techniques (see e.g.~\cite{Qu:2019gqs,ATLAS:2022rkn} and references therein) include all objects associated with the jet and thus their improved performance compared to traditional algorithms could be related at least partially to the dead cone effect. This topic should be investigated further.

\acknowledgments

We would like to thank Valery Khoze for his interest in this work and helpful comments. 

\appendix

\section{Hadron $\xi_p$-distribution at $W=8$~GeV from interpolation}
\label{sec:interpolation}

The $\xi_p$-distributions of hadrons at $W=8$~GeV are obtained by interpolation between the neighbouring energies. Such data have been collected by the BES~\cite{BES:2003xdf} and TASSO~\cite{TASSO:1990cdg} experiments at $W=2-5$~GeV and $14-44$~GeV, respectively, and can be fitted by a Distorted Gaussian, acknowledged to be well suited for QCD analysis~\cite{Fong:1990nt,Perez-Ramos:2013eba}:
\begin{equation}
  \bar D\left(\xi,W\right) = \frac{N}{\sigma\sqrt{2\pi}}\exp{\left[\frac{1}{8}k-\frac{1}{2}s\delta-\frac{1}{4}(2+k)\delta^2+\frac{1}{6}s\delta^3+\frac{1}{24}k\delta^4\right]},
\label{Gauss}
\end{equation}
with $\delta=(\xi-\bar \xi)/\sigma$ and the mean $\bar \xi$. The parameters in eq.~(\ref{Gauss}) are the mean multiplicity $N^{ch}$, mean value $\bar \xi$ or maximum peak position $\xi_{max}=\bar \xi-\frac{1}{2}\sigma s$, width $\sigma$, skewness $s$ and kurtosis $k$. These moments show only a smooth dependence as function of $\ln W$~\cite{Perez-Ramos:2013eba}. Therefore, the first five moments are determined at $W=4.8$ and 14~GeV from a fit to the $\xi_p$ spectra, and those at $W=8.0$~GeV by interpolation in $\ln W$, see Tab.~\ref{tab:interpolation}. 

\begin{table}[htbp!]
\caption{Moments of the Distorted Gaussian distribution (DG) for BES and TASSO data and interpolated DG at $W=8.0$~GeV. The errors for $N$, $\xi_{\rm max}$ and $\sigma$ are $1-4$\%, $1-5$\% and $3-10$\%, respectively, while $s$ and $k$ are not well determined by the available data.} 
\label{tab:interpolation}
\begin{ruledtabular}
\begin{tabular}{lddddd}
Moments & 
\multicolumn{1}{l}{ $N$ } &  
\multicolumn{1}{l}{ $\xi_{\rm max}$ } & 
\multicolumn{1}{l}{ $\sigma$ } &
\multicolumn{1}{l}{ $s$ } & 
\multicolumn{1}{l}{ $k$ } \\
\colrule
BES (4.8 GeV) &  $4.62$  & $1.81$  & $0.70$ &  $0.72$ & $0.14$ \\
Interp. DG (8 GeV) &  $6.49$ & $2.05$ &  $0.74$  & $0.37$ & $-0.20$\\
TASSO (14 GeV) &  $8.73$ & $2.35$ & $0.80$ & $-0.05$ & $-0.60$ \\
\end{tabular}
\end{ruledtabular}
\end{table}

\begin{figure}[htb]
\begin{tabular}{cc}
\hskip -1.1cm
\includegraphics[height=9.0cm,width=9.5cm]{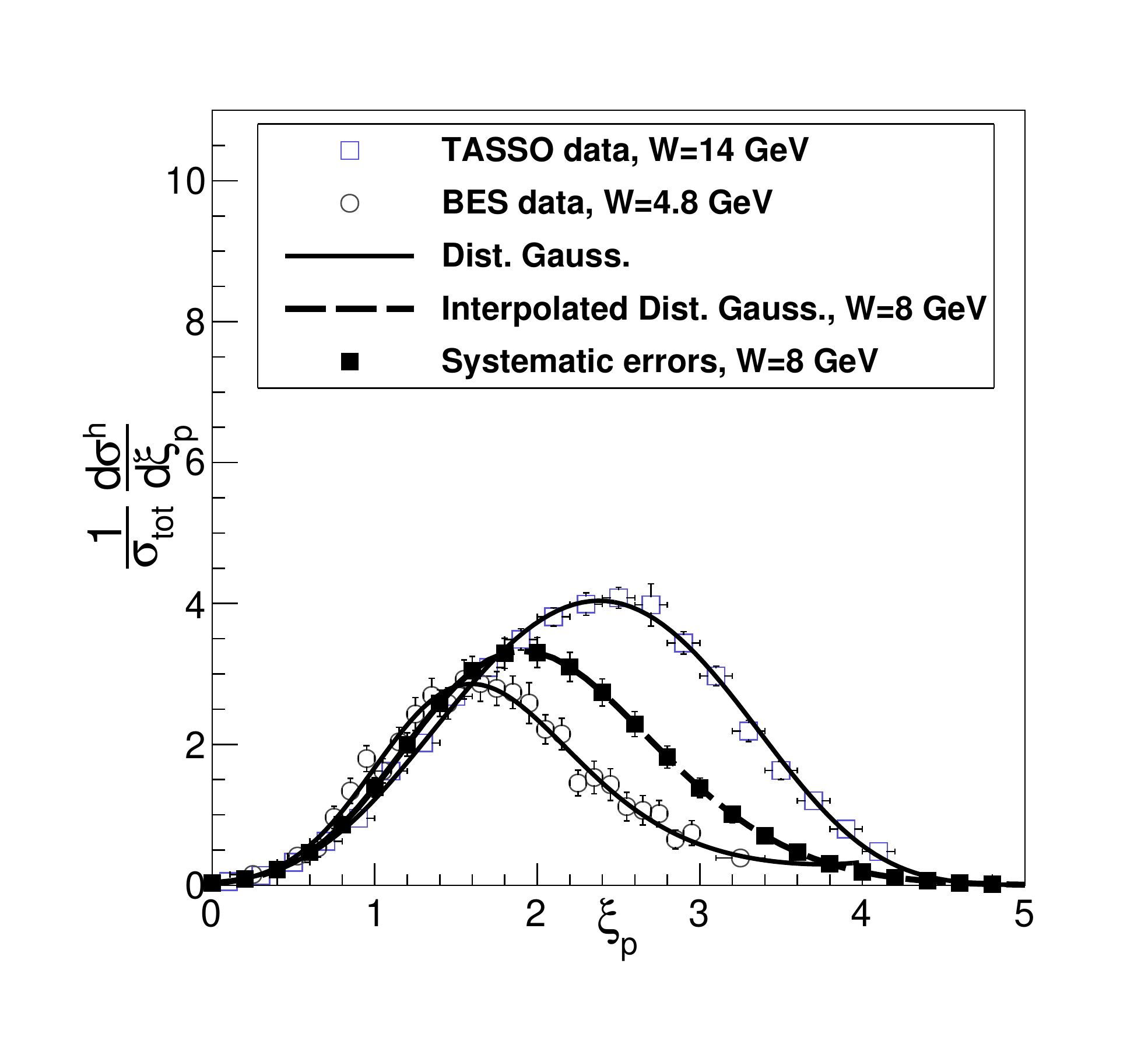} &
\hskip -1.0cm
\includegraphics[height=9.0cm,width=9.5cm]{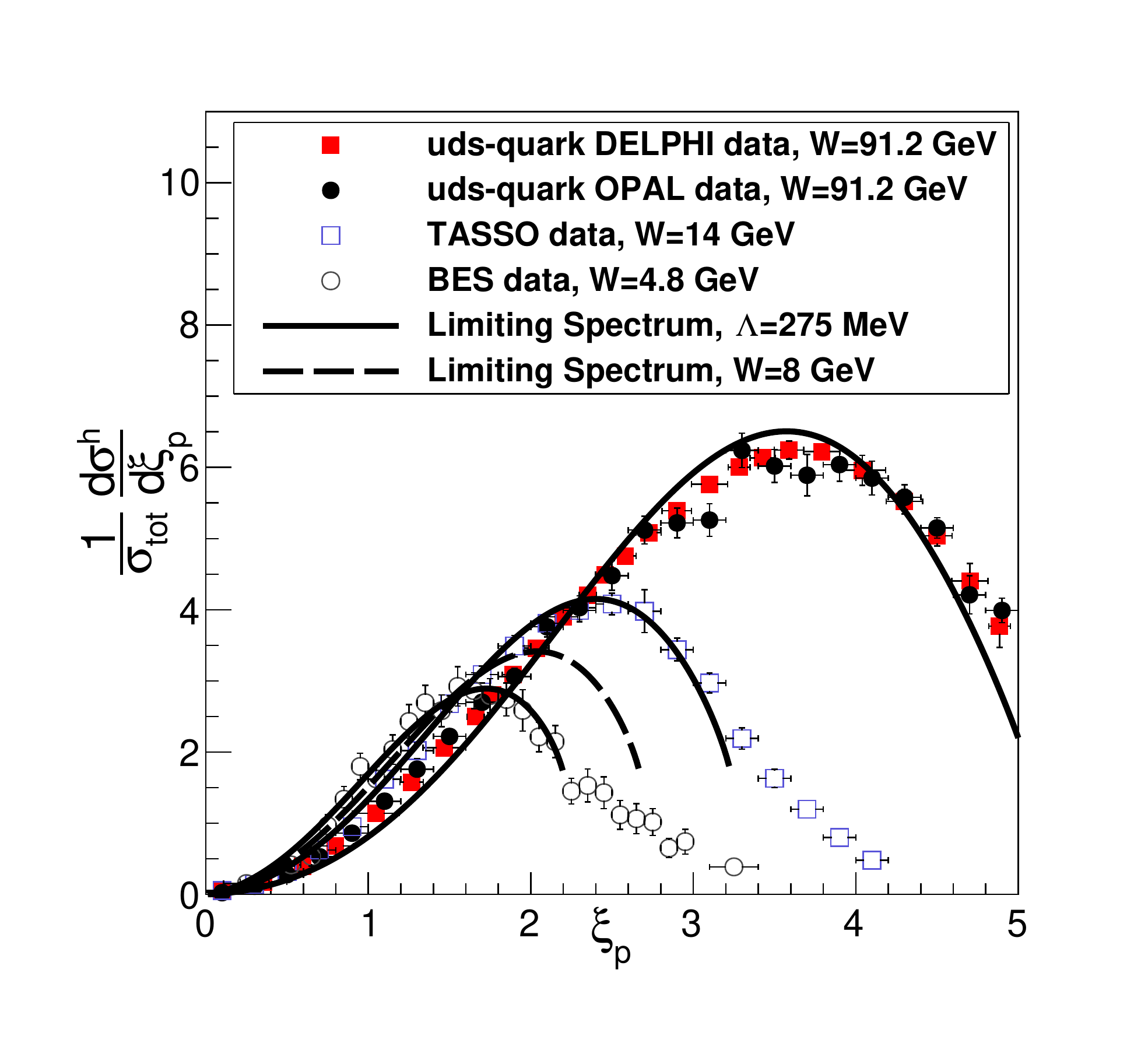} \\
\end{tabular}
\caption{Distribution of charged hadrons in $\xi_p= \ln(1/x_p)$ at c.m.s.\ energies $W=4.8$ and 14~GeV together with Distorted Gaussian fits and the interpolated spectrum at $W=8.0$~GeV with systematic errors (left panel); the same data and the uds-quark data at $W=91.2$~GeV as compared to the MLLA Limiting Spectrum and prediction for $W=8.0$~GeV using $\Lambda_{QCD}=275$~MeV, $K_{ch}=1.28$ at 91.2~GeV and $K_{ch}=1.52$ at lower energies (right panel).}
\label{fig:theoretical_curve}
\end{figure}

In Fig.~\ref{fig:theoretical_curve}, the results of the fits and the interpolated spectrum at $W=8.0$~GeV are displayed. An estimate of the errors is obtained by averaging the relative errors of $\bar D(\xi,W)$ of the BES and TASSO data at their maxima $\delta \bar D_{max}$ (6$\%$), and by increasing their values for smaller $\bar D(\xi)$ as $\delta \bar D(\xi) = \delta \bar D_{max}\sqrt{\bar D_{max}/\bar D(\xi)}$ so as to account for statistical fluctuations.

At $W=8.0$~GeV, there is also some production of c-quarks, which contributes to the multiplicity $N^{ch}=6.49\pm0.19$ with $0.4\pm 0.2$ units according to the estimate in~\cite{Dokshitzer:2005ri}, such that $N^{ch}_{uds}=6.1\pm 0.3$ is taken in the following. In order to obtain the $\xi_p$-distribution for uds-quarks, the distorted Gaussian fit obtained with parameters displayed in Tab.~\ref{tab:interpolation} has to be corrected for charm particle decays. This small correction is approximated by a global lowering of the $\xi_p$-distribution by 6$\%$ and a systematic error of 3$\%$ is linearly added to the experimental error above. 

In Fig.~\ref{fig:theoretical_curve} (right panel), the uds-quark $\xi_p$-distribution data by DELPHI~\cite{DELPHI:1998cgx} and OPAL~\cite{OPAL:1998arz} at $W=91.2$~GeV and the data by BES~\cite{BES:2003xdf} at $W=4.8$~GeV and TASSO~\cite{TASSO:1990cdg} at $W=14$~GeV have been fitted with the Limiting Spectrum with the same $\Lambda=275$~MeV, the normalisation was changed from $K_{ch}=1.28$ at $W=91.2$~GeV to $K_{ch}=1.52$ at the lower energies. There is a good overall description of the data by the ``hump backed plateau'' distribution rising and broadening between 4.8 and 91.2~GeV with deviations of up to 10$\%$ from the data, except for the high end of the distribution, where there is a kinematic limit for the massless partons at $p=Q_0=275$~MeV. With these parameters, the Limiting Spectrum at $W=8.0$~GeV, shown as dashed curve in Fig.~\ref{fig:theoretical_curve} (right panel), can be found. The maximum of this curve is consistent with the value $dn/d\xi|_{max}(8\ \rm{GeV})=3.5\pm0.1$ found by interpolating, as before, the corresponding maxima in the BES~\cite{BES:2003xdf} and the TASSO~\cite{TASSO:1990cdg} data.

The uds-quark $\xi_p$-distribution at $W=8.0$~GeV is obtained again after charm decays are corrected for by rescaling the $\xi_p$ spectrum by 6$\%$ using $K_{ch}=1.43$. 


\begin{thebibliography}{37}%
\makeatletter
\providecommand \@ifxundefined [1]{%
 \@ifx{#1\undefined}
}%
\providecommand \@ifnum [1]{%
 \ifnum #1\expandafter \@firstoftwo
 \else \expandafter \@secondoftwo
 \fi
}%
\providecommand \@ifx [1]{%
 \ifx #1\expandafter \@firstoftwo
 \else \expandafter \@secondoftwo
 \fi
}%
\providecommand \natexlab [1]{#1}%
\providecommand \enquote  [1]{``#1''}%
\providecommand \bibnamefont  [1]{#1}%
\providecommand \bibfnamefont [1]{#1}%
\providecommand \citenamefont [1]{#1}%
\providecommand \href@noop [0]{\@secondoftwo}%
\providecommand \href [0]{\begingroup \@sanitize@url \@href}%
\providecommand \@href[1]{\@@startlink{#1}\@@href}%
\providecommand \@@href[1]{\endgroup#1\@@endlink}%
\providecommand \@sanitize@url [0]{\catcode `\\12\catcode `\$12\catcode
  `\&12\catcode `\#12\catcode `\^12\catcode `\_12\catcode `\%12\relax}%
\providecommand \@@startlink[1]{}%
\providecommand \@@endlink[0]{}%
\providecommand \url  [0]{\begingroup\@sanitize@url \@url }%
\providecommand \@url [1]{\endgroup\@href {#1}{\urlprefix }}%
\providecommand \urlprefix  [0]{URL }%
\providecommand \Eprint [0]{\href }%
\providecommand \doibase [0]{https://doi.org/}%
\providecommand \selectlanguage [0]{\@gobble}%
\providecommand \bibinfo  [0]{\@secondoftwo}%
\providecommand \bibfield  [0]{\@secondoftwo}%
\providecommand \translation [1]{[#1]}%
\providecommand \BibitemOpen [0]{}%
\providecommand \bibitemStop [0]{}%
\providecommand \bibitemNoStop [0]{.\EOS\space}%
\providecommand \EOS [0]{\spacefactor3000\relax}%
\providecommand \BibitemShut  [1]{\csname bibitem#1\endcsname}%
\let\auto@bib@innerbib\@empty
\bibitem [{\citenamefont {Dokshitzer}\ \emph
  {et~al.}(1991{\natexlab{a}})\citenamefont {Dokshitzer}, \citenamefont
  {Khoze},\ and\ \citenamefont {Troian}}]{Dokshitzer:1991fc}%
  \BibitemOpen
  \bibfield  {author} {\bibinfo {author} {\bibfnamefont {Y.~L.}\ \bibnamefont
  {Dokshitzer}}, \bibinfo {author} {\bibfnamefont {V.~A.}\ \bibnamefont
  {Khoze}},\ and\ \bibinfo {author} {\bibfnamefont {S.~I.}\ \bibnamefont
  {Troian}},\ }\bibfield  {title} {\bibinfo {title} {{Particle spectra in light
  and heavy quark jets}},\ }\href {https://doi.org/10.1088/0954-3899/17/10/003}
  {\bibfield  {journal} {\bibinfo  {journal} {J. Phys. G}\ }\textbf {\bibinfo
  {volume} {17}},\ \bibinfo {pages} {1481} (\bibinfo {year}
  {1991}{\natexlab{a}})}\BibitemShut {NoStop}%
\bibitem [{\citenamefont {Dokshitzer}\ \emph
  {et~al.}(1991{\natexlab{b}})\citenamefont {Dokshitzer}, \citenamefont
  {Khoze},\ and\ \citenamefont {Troian}}]{Dokshitzer:1991fd}%
  \BibitemOpen
  \bibfield  {author} {\bibinfo {author} {\bibfnamefont {Y.~L.}\ \bibnamefont
  {Dokshitzer}}, \bibinfo {author} {\bibfnamefont {V.~A.}\ \bibnamefont
  {Khoze}},\ and\ \bibinfo {author} {\bibfnamefont {S.~I.}\ \bibnamefont
  {Troian}},\ }\bibfield  {title} {\bibinfo {title} {{On specific QCD
  properties of heavy quark fragmentation ('dead cone')}},\ }\href
  {https://doi.org/10.1088/0954-3899/17/10/023} {\bibfield  {journal} {\bibinfo
   {journal} {J. Phys. G}\ }\textbf {\bibinfo {volume} {17}},\ \bibinfo {pages}
  {1602} (\bibinfo {year} {1991}{\natexlab{b}})}\BibitemShut {NoStop}%
\bibitem [{\citenamefont {Schumm}\ \emph {et~al.}(1992)\citenamefont {Schumm},
  \citenamefont {Dokshitzer}, \citenamefont {Khoze},\ and\ \citenamefont
  {Koetke}}]{Schumm:1992xt}%
  \BibitemOpen
  \bibfield  {author} {\bibinfo {author} {\bibfnamefont {B.~A.}\ \bibnamefont
  {Schumm}}, \bibinfo {author} {\bibfnamefont {Y.~L.}\ \bibnamefont
  {Dokshitzer}}, \bibinfo {author} {\bibfnamefont {V.~A.}\ \bibnamefont
  {Khoze}},\ and\ \bibinfo {author} {\bibfnamefont {D.~S.}\ \bibnamefont
  {Koetke}},\ }\bibfield  {title} {\bibinfo {title} {{MLLA and the average
  charged multiplicity of events containing heavy quarks in e+ e-
  annihilation}},\ }\href {https://doi.org/10.1103/PhysRevLett.69.3025}
  {\bibfield  {journal} {\bibinfo  {journal} {Phys. Rev. Lett.}\ }\textbf
  {\bibinfo {volume} {69}},\ \bibinfo {pages} {3025} (\bibinfo {year}
  {1992})}\BibitemShut {NoStop}%
\bibitem [{\citenamefont {Dokshitzer}\ \emph {et~al.}(2006)\citenamefont
  {Dokshitzer}, \citenamefont {Fabbri}, \citenamefont {Khoze},\ and\
  \citenamefont {Ochs}}]{Dokshitzer:2005ri}%
  \BibitemOpen
  \bibfield  {author} {\bibinfo {author} {\bibfnamefont {Y.~L.}\ \bibnamefont
  {Dokshitzer}}, \bibinfo {author} {\bibfnamefont {F.}~\bibnamefont {Fabbri}},
  \bibinfo {author} {\bibfnamefont {V.~A.}\ \bibnamefont {Khoze}},\ and\
  \bibinfo {author} {\bibfnamefont {W.}~\bibnamefont {Ochs}},\ }\bibfield
  {title} {\bibinfo {title} {{Multiplicity difference between heavy and light
  quark jets revisited}},\ }\href {https://doi.org/10.1140/epjc/s2005-02424-5}
  {\bibfield  {journal} {\bibinfo  {journal} {Eur. Phys. J. C}\ }\textbf
  {\bibinfo {volume} {45}},\ \bibinfo {pages} {387} (\bibinfo {year} {2006})},\
  \Eprint {https://arxiv.org/abs/hep-ph/0508074} {arXiv:hep-ph/0508074}
  \BibitemShut {NoStop}%
\bibitem [{\citenamefont {Acharya}\ \emph {et~al.}(2022)\citenamefont {Acharya}
  \emph {et~al.}}]{ALICE:2021aqk}%
  \BibitemOpen
  \bibfield  {author} {\bibinfo {author} {\bibfnamefont {S.}~\bibnamefont
  {Acharya}} \emph {et~al.} (\bibinfo {collaboration} {ALICE}),\ }\bibfield
  {title} {\bibinfo {title} {{Direct observation of the dead-cone effect in
  quantum chromodynamics}},\ }\href
  {https://doi.org/10.1038/s41586-022-04572-w} {\bibfield  {journal} {\bibinfo
  {journal} {Nature}\ }\textbf {\bibinfo {volume} {605}},\ \bibinfo {pages}
  {440} (\bibinfo {year} {2022})},\ \bibinfo {note} {[Erratum: Nature 607, E22
  (2022)]},\ \Eprint {https://arxiv.org/abs/2106.05713} {arXiv:2106.05713
  [nucl-ex]} \BibitemShut {NoStop}%
\bibitem [{\citenamefont {Perieanu}(2006)}]{Perieanu:2006vn}%
  \BibitemOpen
  \bibfield  {author} {\bibinfo {author} {\bibfnamefont {A.}~\bibnamefont
  {Perieanu}},\ }\emph {\bibinfo {title} {{The Structure of Charm Jets and the
  Dead Cone Effect in Deep-Inelastic Scattering at HERA}}},\ \href
  {https://doi.org/10.3204/DESY-THESIS-2006-002} {Ph.D. thesis},\ \bibinfo
  {school} {Hamburg U.} (\bibinfo {year} {2006})\BibitemShut {NoStop}%
\bibitem [{\citenamefont {Battaglia}\ \emph {et~al.}(2004)\citenamefont
  {Battaglia}, \citenamefont {Orava},\ and\ \citenamefont
  {Salmi}}]{Battaglia:2004coa}%
  \BibitemOpen
  \bibfield  {author} {\bibinfo {author} {\bibfnamefont {M.}~\bibnamefont
  {Battaglia}}, \bibinfo {author} {\bibfnamefont {R.}~\bibnamefont {Orava}},\
  and\ \bibinfo {author} {\bibfnamefont {L.}~\bibnamefont {Salmi}},\
  }\href@noop {} {\bibinfo {title} {{A Study of depletion of fragmentation
  particles at small angles in b-jets with the DELPHI detector at LEP}}}
  (\bibinfo {year} {2004}),\ \bibinfo {note} {{DELPHI-2004-037 CONF
  712}}\BibitemShut {NoStop}%
\bibitem [{\citenamefont {Bassetto}\ \emph {et~al.}(1983)\citenamefont
  {Bassetto}, \citenamefont {Ciafaloni},\ and\ \citenamefont
  {Marchesini}}]{Bassetto:1983mvz}%
  \BibitemOpen
  \bibfield  {author} {\bibinfo {author} {\bibfnamefont {A.}~\bibnamefont
  {Bassetto}}, \bibinfo {author} {\bibfnamefont {M.}~\bibnamefont
  {Ciafaloni}},\ and\ \bibinfo {author} {\bibfnamefont {G.}~\bibnamefont
  {Marchesini}},\ }\bibfield  {title} {\bibinfo {title} {{Jet Structure and
  Infrared Sensitive Quantities in Perturbative QCD}},\ }\href
  {https://doi.org/10.1016/0370-1573(83)90083-2} {\bibfield  {journal}
  {\bibinfo  {journal} {Phys. Rept.}\ }\textbf {\bibinfo {volume} {100}},\
  \bibinfo {pages} {201} (\bibinfo {year} {1983})}\BibitemShut {NoStop}%
\bibitem [{\citenamefont {Fadin}(1983)}]{Fadin:1983aw}%
  \BibitemOpen
  \bibfield  {author} {\bibinfo {author} {\bibfnamefont {V.~S.}\ \bibnamefont
  {Fadin}},\ }\bibfield  {title} {\bibinfo {title} {{Double logarithmic
  asymptotics of the cross sections of e+ e- annihilation into quarks and
  gluons. (in Russian)}},\ }\href@noop {} {\bibfield  {journal} {\bibinfo
  {journal} {Yad. Fiz.}\ }\textbf {\bibinfo {volume} {37}},\ \bibinfo {pages}
  {408} (\bibinfo {year} {1983})}\BibitemShut {NoStop}%
\bibitem [{\citenamefont {Marchesini}\ and\ \citenamefont
  {Webber}(1983)}]{Marchesini:1983bm}%
  \BibitemOpen
  \bibfield  {author} {\bibinfo {author} {\bibfnamefont {G.}~\bibnamefont
  {Marchesini}}\ and\ \bibinfo {author} {\bibfnamefont {B.~R.}\ \bibnamefont
  {Webber}},\ }\bibfield  {title} {\bibinfo {title} {{Simulation of QCD Jets
  Including Soft Gluon Interference}},\ }\href
  {https://doi.org/10.1016/0550-3213(84)90463-2} {\bibfield  {journal}
  {\bibinfo  {journal} {Nucl. Phys. B}\ }\textbf {\bibinfo {volume} {238}},\
  \bibinfo {pages} {1} (\bibinfo {year} {1983})}\BibitemShut {NoStop}%
\bibitem [{\citenamefont {Dokshitzer}\ \emph
  {et~al.}(1991{\natexlab{c}})\citenamefont {Dokshitzer}, \citenamefont
  {Khoze}, \citenamefont {Mueller},\ and\ \citenamefont {Troian}}]{DKMT-book}%
  \BibitemOpen
  \bibfield  {author} {\bibinfo {author} {\bibfnamefont {Y.~L.}\ \bibnamefont
  {Dokshitzer}}, \bibinfo {author} {\bibfnamefont {V.~A.}\ \bibnamefont
  {Khoze}}, \bibinfo {author} {\bibfnamefont {A.~H.}\ \bibnamefont {Mueller}},\
  and\ \bibinfo {author} {\bibfnamefont {S.~I.}\ \bibnamefont {Troian}},\
  }\href@noop {} {\emph {\bibinfo {title} {Basics of perturbative QCD}}}\
  (\bibinfo  {publisher} {Editions Frontieres},\ \bibinfo {year}
  {1991})\BibitemShut {NoStop}%
\bibitem [{\citenamefont {Azimov}\ \emph {et~al.}(1985)\citenamefont {Azimov},
  \citenamefont {Dokshitzer}, \citenamefont {Khoze},\ and\ \citenamefont
  {Troyan}}]{Azimov:1984np}%
  \BibitemOpen
  \bibfield  {author} {\bibinfo {author} {\bibfnamefont {Y.~I.}\ \bibnamefont
  {Azimov}}, \bibinfo {author} {\bibfnamefont {Y.~L.}\ \bibnamefont
  {Dokshitzer}}, \bibinfo {author} {\bibfnamefont {V.~A.}\ \bibnamefont
  {Khoze}},\ and\ \bibinfo {author} {\bibfnamefont {S.~I.}\ \bibnamefont
  {Troyan}},\ }\bibfield  {title} {\bibinfo {title} {{Similarity of Parton and
  Hadron Spectra in QCD Jets}},\ }\href {https://doi.org/10.1007/BF01642482}
  {\bibfield  {journal} {\bibinfo  {journal} {Z. Phys. C}\ }\textbf {\bibinfo
  {volume} {27}},\ \bibinfo {pages} {65} (\bibinfo {year} {1985})}\BibitemShut
  {NoStop}%
\bibitem [{\citenamefont {Azimov}\ \emph {et~al.}(1986)\citenamefont {Azimov},
  \citenamefont {Dokshitzer}, \citenamefont {Khoze},\ and\ \citenamefont
  {Troyan}}]{Azimov:1985by}%
  \BibitemOpen
  \bibfield  {author} {\bibinfo {author} {\bibfnamefont {Y.~I.}\ \bibnamefont
  {Azimov}}, \bibinfo {author} {\bibfnamefont {Y.~L.}\ \bibnamefont
  {Dokshitzer}}, \bibinfo {author} {\bibfnamefont {V.~A.}\ \bibnamefont
  {Khoze}},\ and\ \bibinfo {author} {\bibfnamefont {S.~I.}\ \bibnamefont
  {Troyan}},\ }\bibfield  {title} {\bibinfo {title} {{Humpbacked QCD Plateau in
  Hadron Spectra}},\ }\href {https://doi.org/10.1007/BF01479529} {\bibfield
  {journal} {\bibinfo  {journal} {Z. Phys. C}\ }\textbf {\bibinfo {volume}
  {31}},\ \bibinfo {pages} {213} (\bibinfo {year} {1986})}\BibitemShut
  {NoStop}%
\bibitem [{\citenamefont {Fong}\ and\ \citenamefont
  {Webber}(1991)}]{Fong:1990nt}%
  \BibitemOpen
  \bibfield  {author} {\bibinfo {author} {\bibfnamefont {C.~P.}\ \bibnamefont
  {Fong}}\ and\ \bibinfo {author} {\bibfnamefont {B.~R.}\ \bibnamefont
  {Webber}},\ }\bibfield  {title} {\bibinfo {title} {{One and two particle
  distributions at small x in QCD jets}},\ }\href
  {https://doi.org/10.1016/0550-3213(91)90302-E} {\bibfield  {journal}
  {\bibinfo  {journal} {Nucl. Phys. B}\ }\textbf {\bibinfo {volume} {355}},\
  \bibinfo {pages} {54} (\bibinfo {year} {1991})}\BibitemShut {NoStop}%
\bibitem [{\citenamefont {Kluth}(2006)}]{Kluth:2006bw}%
  \BibitemOpen
  \bibfield  {author} {\bibinfo {author} {\bibfnamefont {S.}~\bibnamefont
  {Kluth}},\ }\bibfield  {title} {\bibinfo {title} {{Tests of Quantum Chromo
  Dynamics at e+ e- Colliders}},\ }\href
  {https://doi.org/10.1088/0034-4885/69/6/R04} {\bibfield  {journal} {\bibinfo
  {journal} {Rept. Prog. Phys.}\ }\textbf {\bibinfo {volume} {69}},\ \bibinfo
  {pages} {1771} (\bibinfo {year} {2006})},\ \Eprint
  {https://arxiv.org/abs/hep-ex/0603011} {arXiv:hep-ex/0603011} \BibitemShut
  {NoStop}%
\bibitem [{\citenamefont {Perez-Ramos}\ and\ \citenamefont
  {d'Enterria}(2014)}]{Perez-Ramos:2013eba}%
  \BibitemOpen
  \bibfield  {author} {\bibinfo {author} {\bibfnamefont {R.}~\bibnamefont
  {Perez-Ramos}}\ and\ \bibinfo {author} {\bibfnamefont {D.}~\bibnamefont
  {d'Enterria}},\ }\bibfield  {title} {\bibinfo {title} {{Energy evolution of
  the moments of the hadron distribution in QCD jets including NNLL resummation
  and NLO running-coupling corrections}},\ }\href
  {https://doi.org/10.1007/JHEP08(2014)068} {\bibfield  {journal} {\bibinfo
  {journal} {JHEP}\ }\textbf {\bibinfo {volume} {08}},\ \bibinfo {pages}
  {068}},\ \Eprint {https://arxiv.org/abs/1310.8534} {arXiv:1310.8534 [hep-ph]}
  \BibitemShut {NoStop}%
\bibitem [{\citenamefont {Khoze}\ and\ \citenamefont
  {Ochs}(1997)}]{Khoze:1996dn}%
  \BibitemOpen
  \bibfield  {author} {\bibinfo {author} {\bibfnamefont {V.~A.}\ \bibnamefont
  {Khoze}}\ and\ \bibinfo {author} {\bibfnamefont {W.}~\bibnamefont {Ochs}},\
  }\bibfield  {title} {\bibinfo {title} {{Perturbative QCD approach to
  multiparticle production}},\ }\href
  {https://doi.org/10.1142/S0217751X97001638} {\bibfield  {journal} {\bibinfo
  {journal} {Int. J. Mod. Phys. A}\ }\textbf {\bibinfo {volume} {12}},\
  \bibinfo {pages} {2949} (\bibinfo {year} {1997})},\ \Eprint
  {https://arxiv.org/abs/hep-ph/9701421} {arXiv:hep-ph/9701421} \BibitemShut
  {NoStop}%
\bibitem [{\citenamefont {Bierlich}\ \emph {et~al.}(2022)\citenamefont
  {Bierlich} \emph {et~al.}}]{Bierlich:2022pfr}%
  \BibitemOpen
  \bibfield  {author} {\bibinfo {author} {\bibfnamefont {C.}~\bibnamefont
  {Bierlich}} \emph {et~al.},\ }\bibfield  {title} {\bibinfo {title} {{A
  comprehensive guide to the physics and usage of PYTHIA 8.3}},\ }\bibfield
  {journal} {\bibinfo  {journal} {SciPost Phys. Codebases}\ }\textbf {\bibinfo
  {volume} {8}},\ \href {https://doi.org/10.21468/SciPostPhysCodeb.8}
  {10.21468/SciPostPhysCodeb.8} (\bibinfo {year} {2022}),\ \Eprint
  {https://arxiv.org/abs/2203.11601} {arXiv:2203.11601 [hep-ph]} \BibitemShut
  {NoStop}%
\bibitem [{\citenamefont {Skands}\ \emph {et~al.}(2014)\citenamefont {Skands},
  \citenamefont {Carrazza},\ and\ \citenamefont {Rojo}}]{Skands:2014pea}%
  \BibitemOpen
  \bibfield  {author} {\bibinfo {author} {\bibfnamefont {P.}~\bibnamefont
  {Skands}}, \bibinfo {author} {\bibfnamefont {S.}~\bibnamefont {Carrazza}},\
  and\ \bibinfo {author} {\bibfnamefont {J.}~\bibnamefont {Rojo}},\ }\bibfield
  {title} {\bibinfo {title} {{Tuning PYTHIA 8.1: the Monash 2013 Tune}},\
  }\href {https://doi.org/10.1140/epjc/s10052-014-3024-y} {\bibfield  {journal}
  {\bibinfo  {journal} {Eur. Phys. J. C}\ }\textbf {\bibinfo {volume} {74}},\
  \bibinfo {pages} {3024} (\bibinfo {year} {2014})},\ \Eprint
  {https://arxiv.org/abs/1404.5630} {arXiv:1404.5630 [hep-ph]} \BibitemShut
  {NoStop}%
\bibitem [{\citenamefont {Abreu}\ \emph {et~al.}(1998)\citenamefont {Abreu}
  \emph {et~al.}}]{DELPHI:1998cgx}%
  \BibitemOpen
  \bibfield  {author} {\bibinfo {author} {\bibfnamefont {P.}~\bibnamefont
  {Abreu}} \emph {et~al.} (\bibinfo {collaboration} {DELPHI}),\ }\bibfield
  {title} {\bibinfo {title} {{$\pi^\pm$, $K^\pm$, p and anti-p production in Z0
  ---\ensuremath{>} q anti-q, Z0 ---\ensuremath{>} b anti-b, Z$^0$
  ---\ensuremath{>} u anti-u, d anti-d, s anti-s}},\ }\href
  {https://doi.org/10.1007/s100529800989} {\bibfield  {journal} {\bibinfo
  {journal} {Eur. Phys. J. C}\ }\textbf {\bibinfo {volume} {5}},\ \bibinfo
  {pages} {585} (\bibinfo {year} {1998})}\BibitemShut {NoStop}%
\bibitem [{\citenamefont {Ackerstaff}\ \emph {et~al.}(1999)\citenamefont
  {Ackerstaff} \emph {et~al.}}]{OPAL:1998arz}%
  \BibitemOpen
  \bibfield  {author} {\bibinfo {author} {\bibfnamefont {K.}~\bibnamefont
  {Ackerstaff}} \emph {et~al.} (\bibinfo {collaboration} {OPAL}),\ }\bibfield
  {title} {\bibinfo {title} {{Measurements of flavor dependent fragmentation
  functions in Z0 --\ensuremath{>} q anti-q events}},\ }\href
  {https://doi.org/10.1007/s100529901067} {\bibfield  {journal} {\bibinfo
  {journal} {Eur. Phys. J. C}\ }\textbf {\bibinfo {volume} {7}},\ \bibinfo
  {pages} {369} (\bibinfo {year} {1999})},\ \Eprint
  {https://arxiv.org/abs/hep-ex/9807004} {arXiv:hep-ex/9807004} \BibitemShut
  {NoStop}%
\bibitem [{\citenamefont {Achard}\ \emph {et~al.}(2004)\citenamefont {Achard}
  \emph {et~al.}}]{L3:2004cdh}%
  \BibitemOpen
  \bibfield  {author} {\bibinfo {author} {\bibfnamefont {P.}~\bibnamefont
  {Achard}} \emph {et~al.} (\bibinfo {collaboration} {L3}),\ }\bibfield
  {title} {\bibinfo {title} {{Studies of hadronic event structure in $e^{+}
  e^{-}$ annihilation from 30-GeV to 209-GeV with the L3 detector}},\ }\href
  {https://doi.org/10.1016/j.physrep.2004.07.002} {\bibfield  {journal}
  {\bibinfo  {journal} {Phys. Rept.}\ }\textbf {\bibinfo {volume} {399}},\
  \bibinfo {pages} {71} (\bibinfo {year} {2004})},\ \Eprint
  {https://arxiv.org/abs/hep-ex/0406049} {arXiv:hep-ex/0406049} \BibitemShut
  {NoStop}%
\bibitem [{\citenamefont {Buskulic}\ \emph {et~al.}(1995)\citenamefont
  {Buskulic} \emph {et~al.}}]{ALEPH:1995njx}%
  \BibitemOpen
  \bibfield  {author} {\bibinfo {author} {\bibfnamefont {D.}~\bibnamefont
  {Buskulic}} \emph {et~al.} (\bibinfo {collaboration} {ALEPH}),\ }\bibfield
  {title} {\bibinfo {title} {{Measurement of alpha-s from scaling violations in
  fragmentation functions in e+ e- annihilation}},\ }\href
  {https://doi.org/10.1016/0370-2693(95)00917-A} {\bibfield  {journal}
  {\bibinfo  {journal} {Phys. Lett. B}\ }\textbf {\bibinfo {volume} {357}},\
  \bibinfo {pages} {487} (\bibinfo {year} {1995})},\ \bibinfo {note} {[Erratum:
  Phys.Lett.B 364, 247--248 (1995)]}\BibitemShut {NoStop}%
\bibitem [{\citenamefont {Abbaneo}\ \emph {et~al.}(2001)\citenamefont {Abbaneo}
  \emph {et~al.}}]{Bdecaymult}%
  \BibitemOpen
  \bibfield  {author} {\bibinfo {author} {\bibfnamefont {D.}~\bibnamefont
  {Abbaneo}} \emph {et~al.} (\bibinfo {collaboration} {ALEPH, CDF, DELPHI, L3,
  OPAL, SLD}),\ }\href@noop {} {\bibinfo {title} {{Combined results on $b$
  hadron production rates and decay properties}}} (\bibinfo {year} {2001}),\
  \Eprint {https://arxiv.org/abs/hep-ex/0112028} {arXiv:hep-ex/0112028}
  \BibitemShut {NoStop}%
\bibitem [{\citenamefont {Khoze}\ \emph {et~al.}(2001)\citenamefont {Khoze},
  \citenamefont {Ochs},\ and\ \citenamefont {Wosiek}}]{Khoze:2001aa}%
  \BibitemOpen
  \bibfield  {author} {\bibinfo {author} {\bibfnamefont {V.~A.}\ \bibnamefont
  {Khoze}}, \bibinfo {author} {\bibfnamefont {W.}~\bibnamefont {Ochs}},\ and\
  \bibinfo {author} {\bibfnamefont {J.}~\bibnamefont {Wosiek}},\ }\bibfield
  {title} {\bibinfo {title} {Analytical qcd and multiparticle production},\
  }\href@noop {} {\bibfield  {journal} {\bibinfo  {journal} {Handbook of QCD
  (Ioffe Festschrift), ed. M.A. Shifman (World Scientific)}\ } (\bibinfo {year}
  {2001})},\ \Eprint {https://arxiv.org/abs/hep-ph/0009298}
  {arXiv:hep-ph/0009298} \BibitemShut {NoStop}%
\bibitem [{\citenamefont {Dokshitzer}\ \emph {et~al.}(1987)\citenamefont
  {Dokshitzer}, \citenamefont {Khoze},\ and\ \citenamefont
  {Troian}}]{Dokshitzer:1987aa}%
  \BibitemOpen
  \bibfield  {author} {\bibinfo {author} {\bibfnamefont {Y.~L.}\ \bibnamefont
  {Dokshitzer}}, \bibinfo {author} {\bibfnamefont {V.~A.}\ \bibnamefont
  {Khoze}},\ and\ \bibinfo {author} {\bibfnamefont {S.~I.}\ \bibnamefont
  {Troian}},\ }\bibfield  {title} {\bibinfo {title} {New perturbative results
  in hadron jet physics},\ }\href@noop {} {\bibfield  {journal} {\bibinfo
  {journal} {Proc. 6th Int. Conf. on Physics in Collision, ed. M. Derrick
  (World Scientific, Singapore)}\ ,\ \bibinfo {pages} {417}} (\bibinfo {year}
  {1987})}\BibitemShut {NoStop}%
\bibitem [{\citenamefont {Workman}\ \emph {et~al.}(2022)\citenamefont {Workman}
  \emph {et~al.}}]{ParticleDataGroup:2022pth}%
  \BibitemOpen
  \bibfield  {author} {\bibinfo {author} {\bibfnamefont {R.~L.}\ \bibnamefont
  {Workman}} \emph {et~al.} (\bibinfo {collaboration} {Particle Data Group}),\
  }\bibfield  {title} {\bibinfo {title} {{Review of Particle Physics}},\ }\href
  {https://doi.org/10.1093/ptep/ptac097} {\bibfield  {journal} {\bibinfo
  {journal} {PTEP}\ }\textbf {\bibinfo {volume} {2022}},\ \bibinfo {pages}
  {083C01} (\bibinfo {year} {2022})}\BibitemShut {NoStop}%
\bibitem [{\citenamefont {Abdallah}\ \emph {et~al.}(2011)\citenamefont
  {Abdallah} \emph {et~al.}}]{DELPHI:2011aa}%
  \BibitemOpen
  \bibfield  {author} {\bibinfo {author} {\bibfnamefont {J.}~\bibnamefont
  {Abdallah}} \emph {et~al.} (\bibinfo {collaboration} {DELPHI}),\ }\bibfield
  {title} {\bibinfo {title} {{A study of the b-quark fragmentation function
  with the DELPHI detector at LEP I and an averaged distribution obtained at
  the Z Pole}},\ }\href {https://doi.org/10.1140/epjc/s10052-011-1557-x}
  {\bibfield  {journal} {\bibinfo  {journal} {Eur. Phys. J. C}\ }\textbf
  {\bibinfo {volume} {71}},\ \bibinfo {pages} {1557} (\bibinfo {year}
  {2011})},\ \Eprint {https://arxiv.org/abs/1102.4748} {arXiv:1102.4748
  [hep-ex]} \BibitemShut {NoStop}%
\bibitem [{\citenamefont {Baines}\ \emph {et~al.}(2006)\citenamefont {Baines}
  \emph {et~al.}}]{Baines:2006uw}%
  \BibitemOpen
  \bibfield  {author} {\bibinfo {author} {\bibfnamefont {J.}~\bibnamefont
  {Baines}} \emph {et~al.},\ }\href@noop {} {\bibinfo {title} {{Heavy quarks
  (Working Group 3): Summary Report for the HERA-LHC Workshop Proceedings}}}
  (\bibinfo {year} {2006}),\ \Eprint {https://arxiv.org/abs/hep-ph/0601164}
  {arXiv:hep-ph/0601164} \BibitemShut {NoStop}%
\bibitem [{\citenamefont {Dokshitzer}\ \emph {et~al.}(1996)\citenamefont
  {Dokshitzer}, \citenamefont {Khoze},\ and\ \citenamefont
  {Troian}}]{Dokshitzer:1995ev}%
  \BibitemOpen
  \bibfield  {author} {\bibinfo {author} {\bibfnamefont {Y.~L.}\ \bibnamefont
  {Dokshitzer}}, \bibinfo {author} {\bibfnamefont {V.~A.}\ \bibnamefont
  {Khoze}},\ and\ \bibinfo {author} {\bibfnamefont {S.~I.}\ \bibnamefont
  {Troian}},\ }\bibfield  {title} {\bibinfo {title} {{Specific features of
  heavy quark production. LPHD approach to heavy particle spectra}},\ }\href
  {https://doi.org/10.1103/PhysRevD.53.89} {\bibfield  {journal} {\bibinfo
  {journal} {Phys. Rev. D}\ }\textbf {\bibinfo {volume} {53}},\ \bibinfo
  {pages} {89} (\bibinfo {year} {1996})},\ \Eprint
  {https://arxiv.org/abs/hep-ph/9506425} {arXiv:hep-ph/9506425} \BibitemShut
  {NoStop}%
\bibitem [{\citenamefont {Dunwoodie}\ \emph {et~al.}(2004)\citenamefont
  {Dunwoodie} \emph {et~al.}}]{BES:2003xdf}%
  \BibitemOpen
  \bibfield  {author} {\bibinfo {author} {\bibfnamefont {W.}~\bibnamefont
  {Dunwoodie}} \emph {et~al.} (\bibinfo {collaboration} {BES}),\ }\bibfield
  {title} {\bibinfo {title} {{Measurement of inclusive momentum spectra and
  multiplicity distributions of charged particles at S**1/2 sim 2-5 GeV}},\
  }\href {https://doi.org/10.1103/PhysRevD.69.072002} {\bibfield  {journal}
  {\bibinfo  {journal} {Phys. Rev. D}\ }\textbf {\bibinfo {volume} {69}},\
  \bibinfo {pages} {072002} (\bibinfo {year} {2004})},\ \Eprint
  {https://arxiv.org/abs/hep-ex/0306055} {arXiv:hep-ex/0306055} \BibitemShut
  {NoStop}%
\bibitem [{\citenamefont {Akrawy}\ \emph {et~al.}(1990)\citenamefont {Akrawy}
  \emph {et~al.}}]{OPAL:1990vmr}%
  \BibitemOpen
  \bibfield  {author} {\bibinfo {author} {\bibfnamefont {M.~Z.}\ \bibnamefont
  {Akrawy}} \emph {et~al.} (\bibinfo {collaboration} {OPAL}),\ }\bibfield
  {title} {\bibinfo {title} {{A Study of coherence of soft gluons in hadron
  jets}},\ }\href {https://doi.org/10.1016/0370-2693(90)91911-T} {\bibfield
  {journal} {\bibinfo  {journal} {Phys. Lett. B}\ }\textbf {\bibinfo {volume}
  {247}},\ \bibinfo {pages} {617} (\bibinfo {year} {1990})}\BibitemShut
  {NoStop}%
\bibitem [{\citenamefont {Albino}\ \emph {et~al.}(2004)\citenamefont {Albino},
  \citenamefont {Kniehl}, \citenamefont {Kramer},\ and\ \citenamefont
  {Ochs}}]{Albino:2004xa}%
  \BibitemOpen
  \bibfield  {author} {\bibinfo {author} {\bibfnamefont {S.}~\bibnamefont
  {Albino}}, \bibinfo {author} {\bibfnamefont {B.~A.}\ \bibnamefont {Kniehl}},
  \bibinfo {author} {\bibfnamefont {G.}~\bibnamefont {Kramer}},\ and\ \bibinfo
  {author} {\bibfnamefont {W.}~\bibnamefont {Ochs}},\ }\bibfield  {title}
  {\bibinfo {title} {{The Evolution of hadron spectra in the modified leading
  logarithm approximation}},\ }\href
  {https://doi.org/10.1140/epjc/s2004-01875-4} {\bibfield  {journal} {\bibinfo
  {journal} {Eur. Phys. J. C}\ }\textbf {\bibinfo {volume} {36}},\ \bibinfo
  {pages} {49} (\bibinfo {year} {2004})},\ \Eprint
  {https://arxiv.org/abs/hep-ph/0404287} {arXiv:hep-ph/0404287} \BibitemShut
  {NoStop}%
\bibitem [{\citenamefont {Braunschweig}\ \emph {et~al.}(1990)\citenamefont
  {Braunschweig} \emph {et~al.}}]{TASSO:1990cdg}%
  \BibitemOpen
  \bibfield  {author} {\bibinfo {author} {\bibfnamefont {W.}~\bibnamefont
  {Braunschweig}} \emph {et~al.} (\bibinfo {collaboration} {TASSO}),\
  }\bibfield  {title} {\bibinfo {title} {{Global Jet Properties at 14-{GeV} to
  44-{GeV} Center-of-mass Energy in $e^+ e^-$ Annihilation}},\ }\href
  {https://doi.org/10.1007/BF01552339} {\bibfield  {journal} {\bibinfo
  {journal} {Z. Phys. C}\ }\textbf {\bibinfo {volume} {47}},\ \bibinfo {pages}
  {187} (\bibinfo {year} {1990})}\BibitemShut {NoStop}%
\bibitem [{\citenamefont {Barker}(2010)}]{Barker:2010pva}%
  \BibitemOpen
  \bibfield  {author} {\bibinfo {author} {\bibfnamefont {G.~J.}\ \bibnamefont
  {Barker}},\ }\bibfield  {title} {\bibinfo {title} {{Tagging Z0
  ---\ensuremath{>} b anti-b events}},\ }\href
  {https://doi.org/10.1007/978-3-642-05279-8_4} {\bibfield  {journal} {\bibinfo
   {journal} {Springer Tracts Mod. Phys.}\ }\textbf {\bibinfo {volume} {236}},\
  \bibinfo {pages} {57} (\bibinfo {year} {2010})}\BibitemShut {NoStop}%
\bibitem [{\citenamefont {Qu}\ and\ \citenamefont
  {Gouskos}(2020)}]{Qu:2019gqs}%
  \BibitemOpen
  \bibfield  {author} {\bibinfo {author} {\bibfnamefont {H.}~\bibnamefont
  {Qu}}\ and\ \bibinfo {author} {\bibfnamefont {L.}~\bibnamefont {Gouskos}},\
  }\bibfield  {title} {\bibinfo {title} {{ParticleNet: Jet Tagging via Particle
  Clouds}},\ }\href {https://doi.org/10.1103/PhysRevD.101.056019} {\bibfield
  {journal} {\bibinfo  {journal} {Phys. Rev. D}\ }\textbf {\bibinfo {volume}
  {101}},\ \bibinfo {pages} {056019} (\bibinfo {year} {2020})},\ \Eprint
  {https://arxiv.org/abs/1902.08570} {arXiv:1902.08570 [hep-ph]} \BibitemShut
  {NoStop}%
\bibitem [{\citenamefont {Aad}\ \emph {et~al.}(2022)\citenamefont {Aad} \emph
  {et~al.}}]{ATLAS:2022rkn}%
  \BibitemOpen
  \bibfield  {author} {\bibinfo {author} {\bibfnamefont {G.}~\bibnamefont
  {Aad}} \emph {et~al.} (\bibinfo {collaboration} {ATLAS}),\ }\href@noop {}
  {\bibinfo {title} {{Graph Neural Network Jet Flavour Tagging with the ATLAS
  Detector}}} (\bibinfo {year} {2022}),\ \bibinfo {note}
  {{ATL-PHYS-PUB-2022-027}}\BibitemShut {NoStop}%
\end{thebibliography}

%

\end{document}